\documentclass[12pt]{article}
\usepackage{amssymb}

\usepackage{graphicx}
\usepackage{amsmath}
\usepackage{subfigure}
\usepackage{calc}
\usepackage{float}

\input{tcilatex}
\input{tcilatex}

\begin{document}

\title{\textbf{The Quasi-Reversibility Method for the Thermoacoustic Tomography and
a Coefficient Inverse Problem}}
\author{Michael V. Klibanov$^{\ast }$, Andrey V. Kuzhuget$^{\ast },$ \and Sergey I.
Kabanikhin$^{\ast \ast }$, and Dmitriy V.\ Nechaev$^{\bigtriangledown }$ \\
$^{\ast }$Department of Mathematics and Statistics\\
University of North Carolina at Charlotte,\\
Charlotte, NC 28223, USA\\
$^{\ast \ast }$ Sobolev Institute of Mathematics\\
of the Siberian Branch \\
of the Russian Academy of Science\\
Prospect Acad. Koptyuga 2,\\
Novosibirsk, 630090, Russia\\
$^{\bigtriangledown }$ Lavrent'ev Institute of Hydrodynamics \\
of the Siberian Branch \\
of the Russian Academy of Science\\
Prospect Acad. Lavrent'eva 15\\
Novosibirsk, 63090, Russia\\
E-mails: mklibanv@uncc.edu; akuzhuge@uncc.edu; \\
kabanikh@math.nsc.ru; nechaev@hydro.nsc.ru}
\maketitle

\begin{abstract}
An inverse problem of the determination of an initial condition in a
hyperbolic equation from the lateral Cauchy data is considered. This problem
has applications to the thermoacoustic tomography, as well as to linearized
coefficient inverse problems of acoustics and electromagnetics. A new
version of the quasi-reversibility method is described.\ This version
requires a new Lipschitz stability estimate, which is obtained via the
Carleman estimate. Numerical results are presented.
\end{abstract}

\bigskip \textbf{KEY WORDS}: Quasi-reversibility method, Carleman estimate,
numerical results, imaging of sharp peaks

\bigskip \textbf{AMS subject classification:} 65N21, 65D10, 65F10

\section{Introduction}

\bigskip In this paper we propose a new version of the Quasi-Reversibility
Method (QRM) for the inverse problem of the determination of an initial
condition in a hyperbolic equation from the lateral Cauchy data.\ We discuss
applications of this inverse problem to thermoacoustic tomography, as well
as to linearized coefficient inverse problems of acoustics and
electromagnetics. Using the Carleman estimate, we prove convergence of our
version of the QRM.\ We also present numerical results. In particular, we
show that this version of the QRM enables one to image $\delta -$ like
functions, i.e., narrow high peaks.

Let $\Omega \subset \mathbb{R}^{n}$ be a convex domain with a piecewise
smooth boundary $\partial \Omega $ and $2R$ be the diameter of $\Omega
,2R=\max_{x,y\in \Omega }$ $\left| x-y\right| .$ Let $T=const.>R.$ Denote $%
Q_{T}=\Omega \times \left( 0,T\right) .$ Consider the elliptic operator $%
L(x,t)$ of the form
\begin{equation*}
L(x,t)u=\Delta u+\sum\limits_{j=1}^{n}b_{j}\left( x,t\right)
u_{j}+b_{0}\left( x,t\right) u_{t}+c\left( x,t\right) u,
\end{equation*}
where $u_{j}:=\partial _{x_{j}}u.$ We assume that all coefficients of the
operator $L$ belong to $C\left( \overline{Q}_{T}\right) .$ Let the function $%
u\in H^{2}\left( Q_{T}\right) $ be a solution of the hyperbolic equation in
the cylinder $Q_{T},$%
\begin{equation}
u_{tt}=L(x,t)u+F\left( x,t\right) \text{ in }Q_{T},  \tag{1.1}
\end{equation}
$F\in L_{2}\left( Q_{T}\right) $ with initial conditions
\begin{equation}
u\left( x,0\right) =\varphi \left( x\right) ,u_{t}\left( x,0\right) =\psi
\left( x\right) ,\varphi \in H^{1}\left( \Omega \right) ,\psi \in
L_{2}\left( \Omega \right) .  \tag{1.2}
\end{equation}
We consider the following

\textbf{Inverse Problem 1.} Let one of functions $\varphi $ or $\psi $ be
known and another one be unknown. Determine that unknown function assuming
that the following functions $f$ and $g$ are given
\begin{equation}
u\mid _{S_{T}}=f\left( x,t\right) ,\text{ }\frac{\partial u}{\partial \nu }%
\mid _{S_{T}}=g\left( x,t\right) ,\text{ }S_{T}=\partial \Omega \times
\left( 0,T\right) ,  \tag{1.3}
\end{equation}
where $\nu $ is the unit outward normal vector at $\partial \Omega .$ We
call the problem of the determination of the function $\varphi $ the ``$%
\varphi -$problem'' and the problem of the determination of the function $%
\psi $ the ``$\psi -$problem''.

In principle, in the case $T>2R$ one should not assume that one of functions
$\varphi $ or $\psi $ is known.\ This is because for $T>2R$ the following
Lipschitz stability estimate takes place (see [4], [10], [11] and Theorem
2.4.1 in [13])
\begin{equation}
\left\| u\right\| _{H^{1}\left( Q_{T}\right) }\leq C\left( \left\| f\right\|
_{H^{1}\left( S_{T}\right) }+\left\| g\right\| _{L_{2}\left( S_{T}\right)
}+\left\| F\right\| _{L_{2}\left( Q_{T}\right) }\right) .  \tag{1.4}
\end{equation}
Here and below $C$ denotes different positive constants depending only on $%
\Omega ,T$ and $C\left( \overline{Q}_{T}\right) $ norms of coefficients of
the operator $L$. However, since numerical studies for the case of the
finite domain were conducted in previous publications [4], [12], we are
interested here in solving the Inverse Problem 1 in an \emph{unbounded}
domain, which was not done before. This leads us to the case $T>R.$ Namely,
we want to solve an analogue of the Inverse Problem 1 in a quadrant,
assuming that the lateral Cauchy data are given only on parts of two
coordinate axis. We are motivated by the publication [14], where the
Lipschitz stability was proven for an analogue of Inverse Problem 1 for the
case of either a quadrant in $\mathbb{R}^{2}$ or a an octant in $\mathbb{R}%
^{3}$, assuming that one of initial conditions (1.2) is zero, and the second
one has a finite support.

\begin{figure}[tbp]
\begin{center}
\includegraphics[scale=0.5]{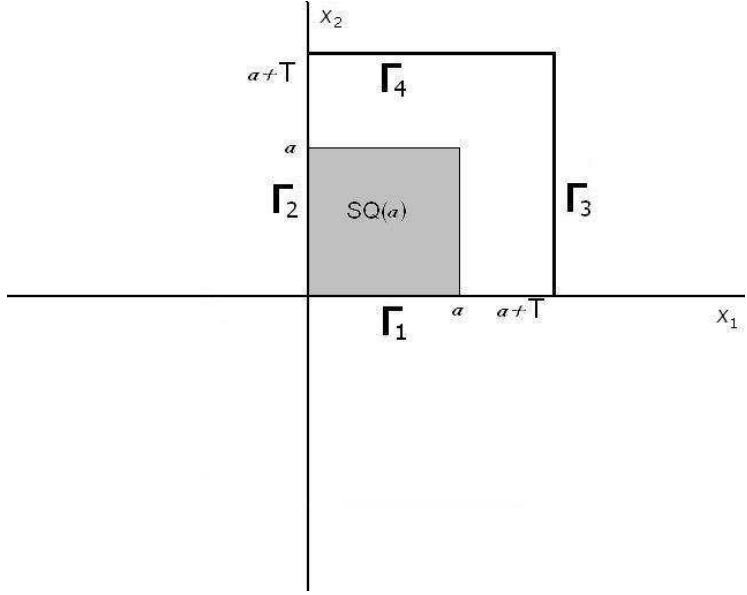}
\end{center}
\caption{Geometry for the Inverse problem 2.}
\label{fig:geometry}
\end{figure}

We now specify conditions of our numerical study. Suppose that equation
(1.1) is homogeneous with $F\left( x,t\right) \equiv 0$ and it is satisfied
in $D_{T}^{3}=\mathbb{R}^{2}\times \left( 0,T\right) .$ Consider the
quadrant $QU=\left\{ x_{1},x_{2}>0\right\} .$ And also consider the square $%
SQ\subset QU,$%
\begin{equation*}
SQ\left( a\right) =\left\{ 0<x_{1},x_{2}<a\right\} .
\end{equation*}
Suppose that
\begin{equation}
\varphi (x)=\psi \left( x\right) =0\text{ outside of }SQ\left( a\right) .
\tag{1.5}
\end{equation}
Then the energy estimate implies that
\begin{equation}
u\left( x,t\right) =0\text{, }\forall \left( x,t\right) \in \left\{ x\mid
x\in QU,dist\left( x,SQ\left( a\right) \right) >T\right\} \times \left(
0,T\right) .  \tag{1.6}
\end{equation}
Denote
\begin{equation*}
\Gamma _{1T}=\left\{ x_{1}\in \left( 0,a+T\right) ,x_{2}=0\right\} \times
\left( 0,T\right) ,
\end{equation*}
\begin{equation*}
\Gamma _{2T}=\left\{ x_{2}\in \left( 0,a+T\right) ,x_{1}=0\right\} \times
\left( 0,T\right) ,
\end{equation*}
\begin{equation*}
\Gamma _{3T}=\left\{ x_{1}=a+T,x_{2}\in \left( 0,a+T\right) \right\} \times
\left( 0,T\right) ,
\end{equation*}
\begin{equation*}
\Gamma _{4T}=\left\{ x_{2}=a+T,x_{1}\in \left( 0,a+T\right) \right\} \times
\left( 0,T\right) ,
\end{equation*}
see Figure $\ref{fig:geometry}$. Then by (1.6)
\begin{equation}
u=\frac{\partial u}{\partial \nu }=0\text{ on }\Gamma _{3T}\cup \Gamma _{4T}.
\tag{1.7}
\end{equation}
Hence, we focus our numerical study on

\textbf{Inverse Problem 2.} Let equation (1.1) be satisfied in $D_{T}^{3}$
with initial conditions (1.2) satisfying (1.4). In this case $\Omega
:=SQ\left( a+T\right) .$Suppose that one of these initial conditions is
zero. Determine the second initial condition, assuming that functions $f$
and $g$ are known, where
\begin{equation}
u\mid _{\Gamma _{1T}\cup \Gamma _{2T}}=f\left( x,t\right) ,\text{ }\frac{%
\partial u}{\partial \nu }\mid _{\Gamma _{1T}\cup \Gamma _{2T}}=g\left(
x,t\right) .  \tag{1.8}
\end{equation}

Suppose for a moment that only the function $f\left( x,t\right) $ is given.\
Then one can solve the boundary value problem for equation (1.1) with $%
F\equiv 0$ outside of the square $SQ\left( a+T\right) $ with the following
initial and boundary data
\begin{equation*}
u\left( x,0\right) =u_{t}\left( x,0\right) =0,x\in R^{2}\diagdown SQ(a+T),
\end{equation*}
\begin{equation*}
u\mid _{\Gamma _{1T}\cup \Gamma _{2T}}=f\left( x,t\right) ,u\mid _{\Gamma
_{3T}\cup \Gamma _{4T}}=0\text{.}
\end{equation*}
This gives one in a \emph{stable way} the normal derivative $g\left(
x,t\right) $ on $\Gamma _{1T}\cup \Gamma _{2T}.$ Thus, we arrive at Inverse
Problem 2. It was proven in [14] that if
\begin{equation}
T>\frac{a\sqrt{2}}{2-\sqrt{2}}  \tag{1.9}
\end{equation}
and one of functions $\varphi $ or $\psi $ equals zero, then the following
Lipschitz stability estimate is valid
\begin{equation}
\left\| u\right\| _{H^{1}\left( G_{T}\right) }\leq C\left( \left\| f\right\|
_{H^{1}\left( \Gamma _{T}\right) }+\left\| g\right\| _{L_{2}\left( \Gamma
_{T}\right) }\right) ,  \tag{1.10}
\end{equation}
where $\Gamma _{T}=\Gamma _{1T}\cup \Gamma _{2T}$ and $\left\| f\right\|
_{H^{1}\left( \Gamma _{T}\right) }=\left\| f\right\| _{H^{1}\left( \Gamma
_{1T}\right) }+\left\| f\right\| _{H^{1}\left( \Gamma _{2T}\right) }.$ The
estimate (1.10) implies a similar estimate for the unknown initial condition
[14]. The knowledge of the fact that one of initial conditions was zero was
used in [14] for either odd or even extension with respect to $t$ of the
function $u(x,t)$ in $\left\{ t<0\right\} ,$ depending on which of initial
conditions was assumed to be unknown$.$ The proof of [14] is based on the
Carleman estimate. The method of Carleman estimates was first applied in
[11] to obtain the Lipschitz stability for the hyperbolic problem with the
lateral Cauchy data, also see [10] and Theorem 2.4.1 in [13]. Prior to [11]
the Lipschitz stability for the hyperbolic problem with the lateral Cauchy
data was obtained in [20] by the method of multipliers, but only for the
case when lower order terms in (1.1) are absent. The use of the Carleman
estimate enables one to incorporate lower order terms and also to extend to
the case of hyperbolic inequalities. Recently the method of [10], [11], [13]
was extended to hyperbolic equations with the non-constant principal part,
see, e.g. [22]-[24]. The method of multipliers was recently extended to the
case of non-zero lower order terms in Theorem 3.5 of the book [6].

The above problems were previously solved numerically in [4], [7], [12] and
[15]. The work [7] was the first one, where the problem of thermoacoustic
tomography was formulated and solved numerically as Inverse Problem 1, i.e.,
as the hyperbolic Cauchy problem with the lateral data. The QRM for the
latter problem was used in [4]. The QRM was first proposed in the book [19]
for a variety of ill-posed boundary value problems. Its convergence rates
were established in [9] and [11] for the cases of Laplace and hyperbolic
equations respectively and in section 2.5 of [13] for elliptic, parabolic
and hyperbolic equations. In particular, it was shown in [13] that one can
work with weak $H^{2}$ solutions of QRM instead of strong $H^{4}$ solutions
of the original book [19]. Also, see [2] and [3] for the recent results for
the QRM for the elliptic case and [8] for the application of the QRM to
linearized coefficient inverse problems for parabolic equations. The main
tool of works [4], [8], [9], [11] and [13] is the tool of Carleman estimates.

There are three main differences between the current paper and the previous
works on the QRM for hyperbolic equations. First, we take into account
boundary conditions via including them in the Tikhonov regularizing
functional $J_{\varepsilon }$. Unlike this, boundary conditions were made
zero in [4] via subtracting off a corresponding function, and they were
treated via integration by parts in [12]. Second, we incorporate in $%
J_{\varepsilon }$ a penalizing term, which reflects our knowledge of one of
initial conditions. We show numerically that without this term we cannot
image well maximal values of the unknown initial condition inside of narrow
peaks. On the other hand, since cancerous tumors can be modeled as narrow
peaks, it is interesting to image those peaks in the application to
thermoacoustic tomography considered below. These first two ideas for $%
J_{\varepsilon }$ are taken from [15]. Mainly because of the second
difference we cannot apply previously derived convergence results and thus,
need to prove convergence of our new version of the QRM. In particular, we
need to prove a \ new Lipschitz stability estimate (Theorem 4.1). Finally,
the third difference is that while $H^{2}$ finite elements were used in [4]
and [12], we use finite differences now. Note that while smooth slowly
varying functions were reconstructed numerically in [4] and [12], our
numerical experiments reconstruct both those functions and $\delta -$like
functions. $\delta -$like functions were also reconstructed in [15] for the
Inverse Problem 2. However, the numerical technique of [15] is different
from one of the current paper. The method of [7] and [15] is based on the
representation of the function $u(x,t)$ via truncated Fourier series and
minimization of the resulting residual least squares functional.

In section 2 we describe applications of Inverse Problems 1 and 2. In
section 3 we describe the version of the QRM we use here. In section 4 we
prove a new Lipschitz stability estimate. In section 5 we prove convergence
of our method, based on the result of section 4. In section 6 we describe
our numerical implementation. In section 7 numerical results are presented.
Conclusions are drawn in section 8.

\section{Applications}

In this section we discuss two applications of above inverse problems

\subsection{Thermoacoustic tomography}

Inverse Problems 1 and 2 arise in thermoacoustic tomography [4], [16], [25].
In \ this case the target is subjected to a short electromagnetic impulse.
The electromagnetic energy is absorbed. As a result, temperature is
increased and the target is expanded. This causes a pressure wave, which is
measured as a change in the acoustic field at the boundary of the sample.
Assuming that the absorption of the electromagnetic energy is spatially
varying inside the sample, the resulting wave field is carrying the
signature of the inhomogeneity. On the other hand, cancerous regions absorb
more than surroundings. This leads to applications in medical imaging.
Hence, the problem is to calculate the absorption coefficient $\alpha (x)$
of the sample using time dependent measurements at its boundary. Let $\beta $
be the thermal expansion coefficient, $c_{p}$ be the specific capacity of
the medium and $I_{0}$ be the power of the source. Usually $\beta ,c_{p}$
and $I_{0}$ are known. Also, assume that the speed of sound in the medium is
constant and equals 1. Let $u(x,t)$ be the pressure wave. It was shown in
e.g., [4] that
\begin{equation}
u_{tt}=\Delta u,\left( x,t\right) \in \mathbb{R}^{3}\times \left( 0,T\right)
,  \tag{2.1}
\end{equation}
\begin{equation}
u\left( x,0\right) =\alpha (x)I_{0}\frac{\beta }{c_{p}},\text{ \ }%
u_{t}\left( x,0\right) =0.  \tag{2.2}
\end{equation}
Suppose that we measure the function $u(x,t)$ at the boundary of the domain $%
\Omega $ and $\alpha (x)=0$ outside of $\Omega .$ Then those measurements
give us the boundary value problem for equation (2.1) outside of $\Omega $
with zero initial conditions. Solving this problem, we uniquely determine
the normal derivative of the function $u(x,t)$ at $\partial \Omega .$ Thus,
we arrive at the $\varphi -$ problem.

A different approach to the problem of thermoacoustic tomography is
currently actively developed in a number of publications. In this approach
the solution of the problem (2.1), (2.2) is presented via the
Poisson-Kirchhoff formula, which leads to the problem of integral geometry
of recovering a function via its integrals over certain spheres, whose
centers run over a surface and radii vary. Then uniqueness theorems are
proven for this case and inversion formulas are derived, see, e,g., [1],
[5], [16], and [17]. In particular, works [1] and [16] include the case of a
variable speed and [16] and [17] include numerical examples. A survey of
these developments can be found in [16]. Also, see \S 1 of Chapter 6 of the
book [18] for an example of the ill-posedness of this integral geometry
problem for the case when centers of spheres run over a plane in $\mathbb{R}%
^{3}$.

\subsection{Linearized inverse acoustic and electromagnetic problems}

There is also another application, in which Inverse Problems 1 and 2 can be
considered as linearized inverse acoustic and inverse electromagnetic
problems. The idea of this subsection is motivated by \S 1 of Chapter 7 of
[18] and \S 3 of Chapter 2 of [21]. We present this application now without
discussing delicate details about the validity of the linearization. In this
setting the point source is running along a surface and time dependent
measurements of back-reflected data are performed at the positions of the
source. In [11] the Newton-Kantorovich method was presented for the case
when the source position is fixed and the time dependent measurements are
performed at a surface.

Let the function $\alpha \left( x\right) \in C\left( \mathbb{R}^{3}\right) $
be strictly positive, $\alpha \left( x\right) \geq const.>0.$ Consider the
Cauchy problem for the hyperbolic equation
\begin{equation}
\alpha \left( x\right) w_{tt}=\Delta _{x}w+4\pi \delta \left(
x-x_{0},t\right) ,\left( x,t\right) \in \mathbb{R}^{3}\times \left(
0,T\right) ,  \tag{2.3}
\end{equation}
\begin{equation}
w\left( x,x_{0},0\right) =w_{t}\left( x,x_{0},0\right) =0,  \tag{2.4}
\end{equation}
where $x_{0}\in \mathbb{R}^{3}$ is the source position. It is well known
that in acoustics $\alpha \left( x\right) =c^{-2}\left( x\right) ,$ where $%
c\left( x\right) $ is the speed of sound in the medium, and in some
situations of the electromagnetics $\alpha \left( x\right) =\left( \mu
\epsilon \right) \left( x\right) ,$ where $\mu $ and $\epsilon $ are
respectively magnetic permeability and electric permittivity of the medium.
We pose the following

\textbf{Inverse Problem 3.} Suppose that the function $\alpha \left(
x\right) =1$ outside of the domain $\Omega $ and it is unknown inside of
this domain. Determine this function for $x\in \Omega ,$ assuming that the
following function $p\left( x_{0},t\right) $ is known
\begin{equation}
w\left( x_{0},x_{0},t\right) \mid _{x_{0}\in \partial \Omega }=p\left(
x_{0},t\right) .  \tag{2.4}
\end{equation}

The full Inverse Problem 3 is difficult to address because of its
nonlinearity. Hence, we consider now a linearized problem. Similarly with \S
3 of Chapter 2 of [21], suppose that the function $\alpha \left( x\right) $
can be represented in the form
\begin{equation*}
\alpha \left( x\right) =1-\xi a\left( x\right) ,
\end{equation*}
where $\xi \in \left( 0,1\right) $ is a small parameter. Hence, the term $%
\xi a\left( x\right) $ is a small perturbation of 1. We assume that this
perturbation is unknown, i.e., the function $a\left( x\right) $ is unknown.
Again, similarly with [21], we can formally set at $\xi \rightarrow 0$
\begin{equation}
w\left( x,x_{0},t\right) =w_{0}\left( x,x_{0},t\right) +\xi w_{1}\left(
x,x_{0},t\right) +O\left( \xi ^{2}\right) ,  \tag{2.5}
\end{equation}
where functions $w_{0}$ and $w_{1}$ are independent on $\xi .$ This setting
was rigorously justified in \S 3 of Chapter 2 of [21] for the case of the
telegraph equation
\begin{equation}
w_{tt}=\Delta w+\left( a_{0}\left( x\right) +\xi a_{1}\left( x\right)
\right) w.  \tag{2.6}
\end{equation}
It was also justified in \S 1 of Chapter 7 of [18]for equation (2.6) without
the introduction of the parameter $\xi ,$ which is actually introduced here
for convenience only. Indeed, instead, one can assume that $\alpha \left(
x\right) =1-a\left( x\right) ,$ where $\left| a\left( x\right) \right| <<1.$

Substituting (2.6) in (2.3) and (2.4) and dropping the term with $O\left(
\xi ^{2}\right) ,$ we obtain that functions $w_{0}$ and $w_{1}$ are
solutions of the following Cauchy problems
\begin{equation}
w_{0tt}=\Delta _{x}w_{0}+4\pi \delta \left( x-x_{0},t\right) ,\left(
x,t\right) \in \mathbb{R}^{3}\times \left( 0,T\right) ,  \tag{2.7}
\end{equation}
\begin{equation}
w_{0}\left( x,x_{0},0\right) =w_{0t}\left( x,x_{0},0\right) =0,  \tag{2.8}
\end{equation}
\begin{equation}
w_{1tt}=\Delta _{x}w_{1}+a\left( x\right) w_{0tt}\left( x,x_{0},t\right)
,\left( x,t\right) \in \mathbb{R}^{3}\times \left( 0,T\right) ,  \tag{2.9}
\end{equation}
\begin{equation}
w_{0}\left( x,x_{0},0\right) =w_{t}\left( x,x_{0},0\right) =0.  \tag{2.10}
\end{equation}
Consider the function $h\left( x,x_{0},t\right) ,$%
\begin{equation*}
h\left( x,x_{0},t\right) =\int\limits_{0}^{t}d\tau \int\limits_{0}^{\tau
}w_{1}\left( x,x_{0},s\right) ds.
\end{equation*}
Integrating (2.9) with respect to $t$ twice and using (2.8) and (2.10), we
obtain
\begin{equation}
h_{tt}=\Delta _{x}h+a\left( x\right) w_{0}\left( x,x_{0},t\right) ,\left(
x,t\right) \in \mathbb{R}^{3}\times \left( 0,T\right) ,  \tag{2.11}
\end{equation}
\begin{equation}
h\left( x,x_{0},0\right) =h_{t}\left( x,x_{0},0\right) =0,  \tag{2.12}
\end{equation}
It follows from (2.7), (2.11), (2.12) and the formula (7.13) of \S 1 of
Chapter 7 of [18] that the function $h\left( x,x_{0},t\right) $ is
\begin{equation}
h\left( x,x_{0},t\right) =\frac{1}{2\pi \left( t^{2}-\left| x-x_{0}\right|
^{2}\right) }\int\limits_{S\left( x,x_{0},t\right) }\left| y-x_{0}\right|
^{2}a\left( y\right) d\omega _{y},  \tag{2.13}
\end{equation}
where $d\omega _{y}=\sin \theta d\varphi d\theta ,\left( \varphi ,\theta
\right) \in \left[ 0,2\pi \right] \times \left[ 0,\pi \right] $ are angles
in the spherical coordinate system with the center at $\left\{ x_{0}\right\}
$ and $S\left( x,x_{0},t\right) $ is the following ellipsoid with foci at $%
\left\{ x\right\} $ and $\left\{ x_{0}\right\} $%
\begin{equation*}
S\left( x,x_{0},t\right) =\left\{ y\in \mathbb{R}^{3}:\left| x-y\right|
+\left| x_{0}-y\right| =t\right\} .
\end{equation*}

Setting in (2.13) $x_{0}:=x$ and denoting $v\left( x,t\right) =h\left(
x,x,t\right) ,$ we obtain that the function $v$ is the spherical Radon
transform of the function $a,$%
\begin{equation}
v\left( x,t\right) =\frac{1}{4\pi }\int\limits_{\left| x-y\right|
=t/2}a\left( y\right) d\omega _{y}.  \tag{2.14}
\end{equation}
On the other hand, (2.14) implies that the function $\widetilde{v}\left(
x,t\right) =v\left( x,2t\right) \cdot t$ is the solution of the following
Cauchy problem
\begin{equation}
\widetilde{v}_{tt}=\Delta \widetilde{v},\left( x,t\right) \in \mathbb{R}%
^{3}\times \left( 0,2T\right) .  \tag{2.15}
\end{equation}
\begin{equation}
\widetilde{v}\mid _{t=0}=0,\widetilde{v}_{t}\mid _{t=0}=a\left( x\right) .
\tag{2.16}
\end{equation}
Also, using the above linearization one can ``translate'' the data $p\left(
x_{0},t\right) $ in (2.4) for the Inverse Problem 3 in the following
function $\widetilde{p}\left( x,t\right) $%
\begin{equation}
\widetilde{v}\mid _{S_{T}}=\widetilde{p}\left( x,t\right) ,t\in \left(
0,2T\right) .  \tag{2.17}
\end{equation}
Since the function $a\left( x\right) =0$ outside of the domain $\Omega ,$
then solving the initial boundary value problem (2.15)-(2.17) for $\left(
x,t\right) \in \left( \mathbb{R}^{3}\diagdown \Omega \right) \times \left(
0,T\right) ,$ we obtain the normal derivative $q\left( x,t\right) ,$%
\begin{equation}
\frac{\partial \widetilde{v}}{\partial \nu }\mid _{S_{T}}=q\left( x,t\right)
,t\in \left( 0,2T\right)  \tag{2.18}
\end{equation}

In conclusion, we have reduced the linearized Inverse Problem 3 to the $\psi
-$problem, which consists in the recovery of the function $a\left( x\right) $
from conditions (2.15)-(2.18). A similar derivation is valid for a similar
inverse problem for the telegraph equation (2.6) at $a_{0}\equiv 0,$ see
[18] and [21].

\section{The Method}

We consider Inverse Problem 1, because it is more general than Inverse
Problem 2. Denote $Mu=u_{tt}-Lu.$ To solve the Inverse Problem 1
numerically, consider the Tikhonov regularizing functional
\begin{equation*}
J_{\varepsilon }\left( u\right) =\left\| Mu-F\right\| _{L_{2}\left(
Q_{T}\right) }^{2}+\varepsilon \left\| u\right\| _{H^{2}\left( Q_{T}\right)
}^{2}
\end{equation*}
\begin{equation}
+\left\| D^{\beta }u\mid _{S_{T}}-D^{\beta }f\right\| _{L_{2}\left(
S_{T}\right) }^{2}+\left\| u_{\nu }\mid _{S_{T}}-g\right\| _{L_{2}\left(
S_{T}\right) }^{2}  \tag{3.1}
\end{equation}
\begin{equation*}
+\chi _{\varphi }\left\| u_{t}(x,0)-\psi \right\| _{L_{2}\left( \Omega
\right) }^{2}+\chi _{\psi }\left\| u(x,0)-\varphi \right\| _{H^{1}\left(
\Omega \right) }^{2},\forall u\in H^{2}\left( Q_{t}\right) .
\end{equation*}
Here $\varepsilon >0$ is the regularization parameter,
\begin{equation*}
u_{\nu }\mid _{S_{T}}:=\frac{\partial u}{\partial \nu }\mid _{S_{T}}
\end{equation*}
and $D^{\beta },\left| \beta \right| \leq 1$ is the operator of $\left(
x,t\right) $ derivatives with$,$ where $x$-derivatives are those, which are
taken in directions orthogonal to the normal vector. Also,
\begin{equation*}
\chi _{\psi }=\left\{
\begin{array}{c}
1\text{ for the }\psi -\text{problem} \\
0\text{ for the }\varphi -\text{problem}
\end{array}
\right\} ,\chi _{\varphi }=\left\{
\begin{array}{c}
1\text{ for the }\varphi -\text{problem} \\
0\text{ for the }\psi -\text{problem}
\end{array}
\right\} .
\end{equation*}
Hence, $\chi _{\varphi }\chi _{\psi }=0,\chi _{\varphi }+\chi _{\psi }=1.$
In previous works on the QRM terms in the second line of (3.1) were absent
because of subtracting off boundary conditions from the original function $u$%
. Terms in the third line of (3.1) were absent also, and they are
incorporated now to emphasize the knowledge of one of initial conditions.

To find the minimizer of $J_{\varepsilon }\left( u\right) ,$ we set the
Fr\'{e}chet derivative of this functional to zero and obtain for all $v\in
H^{2}\left( Q_{T}\right) $
\begin{equation*}
\int\limits_{Q_{T}}MuMvdxdt+\int\limits_{S_{T}}\left( D^{\beta }vD^{\beta
}u+vu\right) \mid _{S_{T}}dS+\int\limits_{S_{T}}\left( v_{\nu }u_{\nu
}\right) \mid _{S_{T}}dS
\end{equation*}
\begin{equation}
+\chi _{\psi }\int\limits_{\Omega }\left[ \nabla u\nabla v+uv\right] \left(
x,0\right) dx+\chi _{\varphi }\int\limits_{\Omega
}u_{t}(x,0)v_{t}(x,0)dx+\varepsilon \left[ u,v\right]  \tag{3.2}
\end{equation}
\begin{equation*}
=\int\limits_{Q_{T}}FMvdxdt+\int\limits_{S_{T}}\sum\limits_{\left| \beta
\right| \leq 1}\left( D^{\beta }v\mid _{S_{T}}\right) D^{\beta
}fdS++\int\limits_{S_{T}}\left( v_{\nu }\mid _{S_{T}}\right) \cdot gdS
\end{equation*}
\begin{equation*}
+\chi _{\psi }\int\limits_{\Omega }\left[ \nabla \varphi \nabla v\left(
x,0\right) +\varphi v\left( x,0\right) \right] dx+\chi _{\varphi
}\int\limits_{\Omega }\psi v_{t}(x,0)dx.
\end{equation*}
Riesz theorem and (3.2) imply

\textbf{Lemma 3.1.} \emph{For any vector function }$\left( F,f,g\right) \in
L_{2}\left( Q_{T}\right) \times H^{1}\left( S_{T}\right) \times L_{2}\left(
S_{T}\right) $\emph{\ there exists unique solution }$u_{\varepsilon }\in
H^{2}\left( Q_{T}\right) $\emph{\ of the problem (3.2) and}
\begin{equation*}
\left\| u_{\varepsilon }\right\| _{H^{2}\left( Q_{T}\right) }\leq \frac{C}{%
\sqrt{\varepsilon }}\left( \left\| F\right\| _{L_{2}\left( Q_{T}\right)
}+\left\| f\right\| _{H^{1}\left( S_{T}\right) }+\left\| g\right\|
_{L_{2}\left( S_{T}\right) }+\chi _{\psi }\left\| \varphi \right\|
_{H^{1}\left( \Omega \right) }+\chi _{\varphi }\left\| \psi \right\|
_{L_{2}\left( \Omega \right) }\right) .
\end{equation*}

Setting in (3.2) $v:=u,$ we obtain that the unique minimizer of the
functional $J_{\varepsilon }\left( u\right) $ satisfies the following
estimate
\begin{equation*}
\left\| Mu\right\| _{L_{2}\left( Q_{T}\right) }^{2}+\chi _{\psi }\left\|
u(x,0)\right\| _{H^{1}\left( \Omega \right) }^{2}+\chi _{\varphi }\left\|
u_{t}(x,0)\right\| _{L_{2}\left( \Omega \right) }^{2}
\end{equation*}
\begin{equation}
+\left\| u\mid _{S_{T}}\right\| _{H^{1}\left( S_{T}\right) }^{2}+\left\|
u_{\nu }\mid _{S_{T}}\right\| _{L_{2}\left( S_{T}\right) }^{2}  \tag{3.3}
\end{equation}
\begin{equation*}
\leq \left\| F\right\| _{L_{2}\left( Q_{T}\right) }^{2}+\left\| f\right\|
_{H^{1}\left( S_{T}\right) }^{2}+\left\| g\right\| _{L_{2}\left(
S_{T}\right) }^{2}+\chi _{\psi }\left\| \varphi \right\| _{H^{1}\left(
\Omega \right) }^{2}+\chi _{\varphi }\left\| \psi \right\| _{H^{1}\left(
\Omega \right) }^{2}.
\end{equation*}
To prove convergence of our method, we need to derive from (3.3) the
Lipschitz stability estimate for the function $u$ in the $H^{1}\left(
Q_{T}\right) $-norm. This in turn requires a modification of the proofs of
[4], [10], [11] and [13]. We specifically refer to the proofs of Theorem
2.4.1 in [13] and Theorem 4.4 in [4]. The main difference with previous
proofs is that now either $\left\| u(x,0)\right\| _{H^{1}\left( \Omega
\right) }^{2}$ or $\left\| u_{t}(x,0)\right\| _{L_{2}\left( \Omega \right)
}^{2}$ can be estimated via the right hand side of (3.3), which was not done
before. This is because terms in the third line of (3.1) were not included
in the Tikhonov functional for QRM. So, if $\chi _{\psi }=0,$ then $\chi
_{\varphi }=1$ and we estimate $\left\| u(x,0)\right\| _{L_{2}\left( \Omega
\right) }^{2}$ in the $\varphi -$problem. If, however, $\chi _{\psi }=1,$
then $\chi _{\varphi }=0$ and we estimate $\left\| u_{t}(x,0)\right\|
_{L_{2}\left( \Omega \right) }^{2}$ in the $\psi $ problem. It is because of
the incorporation of these terms that we can assume that $T>R$, as it is
required in Inverse Problem 2 (see (1.9)), instead of $T>2R$ of previous
works. The required modification is done in the next section.

\section{Lipschitz Stability Estimate}

\textbf{Theorem 4.1.} \emph{Let }$\Omega \subset \mathbb{R}^{n}$\emph{\ be a
convex bounded domain with the piecewise smooth boundary\ and let }$T>R$%
\emph{. Suppose that the function }$u\in H^{2}\left( Q_{T}\right) $\emph{\
satisfies the inequality}
\begin{equation}
\left\| Mu\right\| _{L_{2}\left( Q_{T}\right) }+\chi _{\psi }\left\|
u(x,0)\right\| _{H^{1}\left( \Omega \right) }+\chi _{\varphi }\left\|
u_{t}(x,0)\right\| _{L_{2}\left( \Omega \right) }  \tag{4.1}
\end{equation}
\begin{equation*}
+\left\| u\mid _{S_{T}}\right\| _{H^{1}\left( S_{T}\right) }+\left\| u_{\nu
}\mid _{S_{T}}\right\| _{L_{2}\left( S_{T}\right) }\leq K,
\end{equation*}
\emph{where }$K=const.>0$\emph{. Then }
\begin{equation}
\left\| u\right\| _{H^{1}\left( Q_{T}\right) }+\chi _{\varphi }\left\|
u\left( x,0\right) \right\| _{H^{1}\left( \Omega \right) }+\chi _{\psi
}\left\| u_{t}\left( x,0\right) \right\| _{L_{2}\left( \Omega \right) }\leq
CK.  \tag{4.2}
\end{equation}

\textbf{Proof}. Choose a pair of points $x^{\prime }$,$y^{\prime }\in
\partial \Omega $ such that $\left| x^{\prime }-y^{\prime }\right| =2R.$ Put
the origin at the point $\left( x+y\right) /2.$ Choose a constant $\eta \in
\left( 0,1\right) $. Consider the function $p\left( x,t\right) ,$
\begin{equation*}
p\left( x,t\right) =\left| x\right| ^{2}-\eta t^{2}.
\end{equation*}
Consider the Carleman Weight Function (CWF) $\mathcal{C}(x,t),$%
\begin{equation*}
\mathcal{C}(x,t)=\exp \left( 2\lambda p\left( x,t\right) \right) ,
\end{equation*}
where $\lambda >1$ is a parameter. For any positive number $b$ denote

$G_{b}=\left\{ \left( x,t\right) \mid p\left( x,t\right) >b,x\in \Omega
,t>0\right\} .$ Choose a sufficiently small number $c\in \left(
0,R^{2}\right) .$ Since $T>R$, then in $G_{c}$%
\begin{equation*}
t^{2}<\frac{R^{2}-c}{\eta }<T^{2},\forall \eta \in \left( \eta _{0},1\right)
,
\end{equation*}
where $\eta _{0}=\eta _{0}\left( T,R\right) \in \left( 0,1\right) .$ Hence, $%
G_{c}\subset Q_{T}.$ \ Choose $\delta \in \left( 0,c\right) $ so small that $%
G_{c+4\delta }\neq \varnothing .$ Note that
\begin{equation}
G_{c+4\delta }\subset G_{c+3\delta }\subset G_{c+2\delta }\subset
G_{c+\delta }\subset G_{c}.  \tag{4.3}
\end{equation}

Denote $M_{0}u=u_{tt}-\Delta u.$ The following pointwise Carleman estimate
takes place
\begin{equation}
\left( M_{0}u\right) ^{2}\mathcal{C}^{2}\geq C\lambda \left( \left| \nabla
_{x,t}u\right| ^{2}+\lambda ^{2}u^{2}\right) \mathcal{C}+\nabla \cdot U+V_{t}%
\text{ in }G_{c},\forall u\in C^{2}\left( \overline{G}_{c}\right) ,\forall
\lambda >\lambda _{0},  \tag{4.4}
\end{equation}
where
\begin{equation}
\left| U\right| +\left| V\right| \leq C\lambda \left( \left| \nabla
_{x,t}u\right| ^{2}+\lambda ^{2}u^{2}\right) \mathcal{C}\text{ in }G_{c}
\tag{4.5}
\end{equation}
and $\lambda _{0}\left( G_{c},\eta \right) >1$ is sufficiently large. In
addition, the function $V$ is estimated as
\begin{equation}
\left| V\right| \leq C\lambda ^{3}t\left( \left| \nabla _{x,t}u\right|
^{2}+u^{2}\right) \mathcal{C}+C\lambda ^{3}\left| u_{t}\right| \left( \left|
\nabla u\right| +\left| u\right| \right) \mathcal{C}\text{ in }G_{c}.
\tag{4.6}
\end{equation}
This Carleman estimate was proven in Theorem 2.2.4 of [13]. It was derived
earlier in \S 4 of Chapter 4 of [18].

Consider the cut-off function $\rho \left( x,t\right) \in C^{2}\left(
\overline{G}_{c}\right) $ such that
\begin{equation}
\rho \left( x,t\right) =\left\{
\begin{array}{c}
1\text{ in }G_{c+2\delta } \\
0\text{ in }G_{c}\diagdown G_{c+\delta } \\
\text{ between 0 and 1 otherwise}
\end{array}
\right\} .  \tag{4.7}
\end{equation}
The existence of such functions is well known. For an arbitrary function $%
u\in C^{2}\left( \overline{G}_{c}\right) $ denote $v=v(u):=\rho u.$ Using
(4.4)-(4.6), we obtain
\begin{equation*}
\int\limits_{G_{c}}\left( M_{0}v\right) ^{2}\mathcal{C}dxdt\geq C\lambda
\int\limits_{G_{c}}\left( \left| \nabla _{x,t}v\right| ^{2}+\lambda
^{2}v^{2}\right) \mathcal{C}dxdt
\end{equation*}
\begin{equation*}
-C\lambda ^{3}\int\limits_{\Omega }\left[ \left| u_{t}\right| \left( \left|
\nabla u\right| +\left| u\right| \right) \right] \left( x,0\right) \exp
\left( 2\lambda \left| x\right| ^{2}\right) dx-C\lambda \int\limits_{S_{T}}%
\left[ \left( D^{\beta }u\right) ^{2}+\lambda ^{2}u_{\nu }^{2}\right]
\mathcal{C}dS.
\end{equation*}
Because by (4.3) $G_{c+2\delta }\subset G_{c},$ then (4.7) implies that the
last inequality can be rewritten as
\begin{equation}
\int\limits_{G_{c}}\left( M_{0}v\right) ^{2}\mathcal{C}dxdt\geq C\lambda
\int\limits_{G_{c+2\delta }}\left( \left| \nabla _{x,t}u\right| ^{2}+\lambda
^{2}u^{2}\right) \mathcal{C}dxdt  \tag{4.8}
\end{equation}
\begin{equation*}
-C\lambda ^{3}\int\limits_{\Omega }\left[ \left| u_{t}\right| \left( \left|
\nabla u\right| +\left| u\right| \right) \right] \left( x,0\right) \exp
\left( 2\lambda \left| x\right| ^{2}\right) dx-C\lambda \int\limits_{S_{T}}%
\left[ \left( D^{\beta }u\right) ^{2}+u_{\nu }^{2}\right] \mathcal{C}dS.
\end{equation*}
By (4.7) the left hand side of (4.8) can be estimated from the above as
\begin{equation*}
\int\limits_{G_{c}}\left( M_{0}v\right) ^{2}\mathcal{C}dxdt\leq
\int\limits_{G_{c+2\delta }}\left( M_{0}u\right) ^{2}\mathcal{C}%
dxdt+C\int\limits_{G_{c}\diagdown G_{c+2\delta }}\left( \left| \nabla
_{x,t}u\right| ^{2}+u^{2}\right) \mathcal{C}dxdt
\end{equation*}
\begin{equation*}
\leq \int\limits_{G_{c+2\delta }}\left( M_{0}u\right) ^{2}\mathcal{C}%
dxdt+C\lambda ^{3}\left\| u\right\| _{H^{1}\left( Q_{T}\right) }^{2}\exp
\left[ 2\lambda \left( c+2\delta \right) \right]
\end{equation*}
\begin{equation*}
\leq \int\limits_{G_{c+2\delta }}\left( Mu\right) ^{2}\mathcal{C}%
dxdt+C\int\limits_{G_{c+2\delta }}\left( \left| \nabla _{x,t}u\right|
^{2}+u^{2}\right) \mathcal{C}dxdt+C\left\| u\right\| _{H^{1}\left(
Q_{T}\right) }\exp \left[ 2\lambda \left( c+2\delta \right) \right] .
\end{equation*}
Substituting this in (4.8), recalling that $\lambda $ is sufficiently large
and using (4.3), we obtain
\begin{equation*}
\int\limits_{G_{c+2\delta }}\left( Mu\right) ^{2}\mathcal{C}dxdt+\lambda
^{3}\left\| u\right\| _{H^{1}\left( Q_{T}\right) }^{2}\exp \left[ 2\lambda
\left( c+2\delta \right) \right]
\end{equation*}
\begin{equation*}
+\lambda \int\limits_{S_{T}}\left[ \left( D^{\beta }u\right) ^{2}+\lambda
^{2}u_{\nu }^{2}\right] \mathcal{C}dS+\lambda ^{3}\int\limits_{\Omega }\left[
\left| u_{t}\right| \left( \left| \nabla u\right| +\left| u\right| \right)
\right] \left( x,0\right) \exp \left( 2\lambda \left| x\right| ^{2}\right) dx
\end{equation*}
\begin{equation*}
\geq C\lambda \int\limits_{G_{c+2\delta }}\left( \left| \nabla
_{x,t}u\right| ^{2}+\lambda ^{2}u^{2}\right) \mathcal{C}dxdt\geq C\lambda
^{3}\exp \left[ 2\lambda \left( c+3\delta \right) \right] \int%
\limits_{G_{c+3\delta }}\left( \left| \nabla _{x,t}u\right|
^{2}+u^{2}\right) dxdt.
\end{equation*}
Let $m=\max_{\overline{G}_{c}}p(x,t).$ Dividing the last inequality by $%
C\lambda ^{3}\exp \left[ 2\lambda \left( c+3\delta \right) \right] ,$ we
obtain
\begin{equation*}
\int\limits_{G_{c+3\delta }}\left( \left| \nabla _{x,t}u\right|
^{2}+u^{2}\right) dxdt\leq
\end{equation*}
\begin{equation}
Ce^{2\lambda m}\left( \left\| Mu\right\| _{L_{2}\left( Q_{T}\right)
}^{2}+\left\| u\mid _{S_{T}}\right\| _{H^{1}\left( S_{T}\right)
}^{2}+\left\| u_{\nu }\mid _{S_{T}}\right\| _{L_{2}\left( S_{T}\right)
}^{2}\right) +C\left\| u\right\| _{H^{1}\left( Q_{T}\right)
}^{2}e^{-2\lambda \delta }  \tag{4.9}
\end{equation}
\begin{equation*}
+Ce^{2\lambda m}\int\limits_{\Omega }\left[ \left| u_{t}\right| \left(
\left| \nabla u\right| +\left| u\right| \right) \right] \left( x,0\right) dx.
\end{equation*}

The last term of (4.9) was not present in previous publications, and we will
analyze it now. Consider the $\varphi -$problem first. That is, consider the
case $\chi _{\varphi }=1,\chi _{\psi }=0.$ Let $\gamma >0$ be a small number
which we will choose later. We estimate the last term of (4.9) as
\begin{equation*}
Ce^{2\lambda m}\int\limits_{\Omega }\left[ \left| u_{t}\right| \left( \left|
\nabla u\right| +\left| u\right| \right) \right] \left( x,0\right) dx
\end{equation*}
\begin{equation}
\leq C\gamma \int\limits_{\Omega }\left( \left| \nabla u\right|
^{2}+u^{2}\right) \left( x,0\right) dx+\frac{Ce^{4\lambda m}}{\gamma }%
\int\limits_{\Omega }u_{t}^{2}\left( x,0\right) dx  \tag{4.10$\varphi $}
\end{equation}
\begin{equation*}
\leq C\gamma \left\| u\left( x,0\right) \right\| _{H^{1}\left( \Omega
\right) }^{2}+\frac{Ce^{4\lambda m}}{\gamma }K^{2}.
\end{equation*}
We have used (4.1) to estimate the last term in the second line of (4.10$%
\varphi $). Consider now the $\psi -$problem. Then similarly with (4.10$%
\varphi $)
\begin{equation}
Ce^{2\lambda m}\int\limits_{\Omega }\left[ \left| u_{t}\right| \left( \left|
\nabla u\right| +\left| u\right| \right) \right] \left( x,0\right) dx\leq
C\gamma \left\| u_{t}\left( x,0\right) \right\| _{L_{2}\left( \Omega \right)
}^{2}+\frac{Ce^{4\lambda m}}{\gamma }K^{2}.  \tag{4.10$\psi $}
\end{equation}
Consider now the set
\begin{equation*}
F_{1}\left( c,\delta \right) =G_{c+3\delta }\cap \left\{ t\in \left(
0,\delta \right) \right\} .
\end{equation*}
Then
\begin{equation}
\left\{ \left( x,t\right) :\left| x\right| >\sqrt{c+3\delta +\eta \delta ^{2}%
},x\in \Omega ,t\in \left( 0,\delta \right) \right\} \subset F_{1}\left(
c,\delta \right) .  \tag{4.11}
\end{equation}
Then (4.9), (4.10$\varphi )$ and (4.10$\psi )$ imply that
\begin{equation*}
\int\limits_{F_{1}\left( c,\delta \right) }\left( \left| \nabla
_{x,t}u\right| ^{2}+u^{2}\right) dxdt\leq \frac{Ce^{4\lambda m}}{\gamma }%
K^{2}+C\left\| u\right\| _{H^{1}\left( Q_{T}\right) }^{2}e^{-2\lambda \delta
}
\end{equation*}
\begin{equation*}
+Ce^{2\lambda m}\left( \left\| u\mid _{S_{T}}\right\| _{H^{1}\left(
S_{T}\right) }^{2}+\left\| u_{\nu }\mid _{S_{T}}\right\| _{L_{2}\left(
S_{T}\right) }^{2}\right)
\end{equation*}
\begin{equation*}
+\chi _{\varphi }C\gamma \left\| u\left( x,0\right) \right\| _{H^{1}\left(
\Omega \right) }^{2}+\chi _{\psi }C\gamma \left\| u_{t}\left( x,0\right)
\right\| _{L_{2}\left( \Omega \right) }^{2}.
\end{equation*}
Hence, by (4.1)
\begin{equation}
\int\limits_{F_{1}\left( c,\delta \right) }\left( \left| \nabla
_{x,t}u\right| ^{2}+u^{2}\right) dxdt\leq \frac{Ce^{4\lambda m}}{\gamma }%
K^{2}+C\left\| u\right\| _{H^{1}\left( Q_{T}\right) }^{2}e^{-2\lambda \delta
}  \tag{4.12}
\end{equation}
\begin{equation*}
+\chi _{\varphi }C\gamma \left\| u\left( x,0\right) \right\| _{H^{1}\left(
\Omega \right) }^{2}+\chi _{\psi }C\gamma \left\| u_{t}\left( x,0\right)
\right\| _{L_{2}\left( \Omega \right) }^{2}.
\end{equation*}

Choose numbers $c$ and $\delta $ so small that $3\sqrt{c+3\delta +\eta
\delta ^{2}}<R.$ Hence, we can choose $x_{0}\in \Omega $ such that $\left|
x_{0}\right| =3\sqrt{c+3\delta +\eta \delta ^{2}}.$\ Next, we ``shift'' the
function $p\left( x,t\right) $ to the point $x_{0},$ thus considering the
function
\begin{equation*}
p\left( x-x_{0},t\right) =\left| x-x_{0}\right| ^{2}-\eta t^{2}.
\end{equation*}
For $b>0$ let $G_{b}\left( x_{0}\right) =\left\{ \left( x,t\right) \mid
p\left( x-x_{0},t\right) >b,x\in \Omega ,t>0\right\} .$ Similarly with the
above denote
\begin{equation*}
F_{2}\left( c,\delta \right) =G_{c+3\delta }\left( x_{0}\right) \cap \left\{
t\in \left( 0,\delta \right) \right\} .
\end{equation*}
Then
\begin{equation}
\left\{ \left( x,t\right) :\left| x-x_{0}\right| >\sqrt{c+3\delta +\eta
\delta ^{2}},x\in \Omega ,t\in \left( 0,\delta \right) \right\} \subset
F_{2}\left( c,\delta \right) .  \tag{4.13}
\end{equation}
Consider an arbitrary point $x$ such that $x\in \left\{ \left| x\right| <2%
\sqrt{c+3\delta +\eta \delta ^{2}}\right\} .$ Then
\begin{equation*}
\left| x_{0}-x\right| \geq \left| x_{0}\right| -\left| x\right| \geq 3\sqrt{%
c+3\delta +\eta \delta ^{2}}-2\sqrt{c+3\delta +\eta \delta ^{2}}=\sqrt{%
c+3\delta +\eta \delta ^{2}}.
\end{equation*}
Hence, by (4.13)
\begin{equation*}
\left\{ \left( x,t\right) :\left| x\right| <2\sqrt{c+3\delta +\eta \delta
^{2}},t\in \left( 0,\delta \right) \right\} \subset F_{2}\left( c,\delta
\right) .
\end{equation*}
Combining this with (4.11), we see that
\begin{equation}
\Omega \times \left( 0,\delta \right) =Q_{\delta }\subset F_{1}\left(
c,\delta \right) \cup F_{2}\left( c,\delta \right) .  \tag{4.14}
\end{equation}
Using function $p\left( x-x_{0},t\right) $ instead of $p\left( x,t\right) ,$
we obtain similarly with (4.12)
\begin{equation*}
\int\limits_{F_{2}\left( c,\delta \right) }\left( \left| \nabla
_{x,t}u\right| ^{2}+u^{2}\right) dxdt\leq \frac{Ce^{4\lambda m}}{\gamma }%
K^{2}+C\left\| u\right\| _{H^{1}\left( Q_{T}\right) }^{2}e^{-2\lambda \delta
}
\end{equation*}
\begin{equation*}
+\chi _{\varphi }C\gamma \left\| u\left( x,0\right) \right\| _{H^{1}\left(
\Omega \right) }^{2}+\chi _{\psi }C\gamma \left\| u_{t}\left( x,0\right)
\right\| _{L_{2}\left( \Omega \right) }^{2}.
\end{equation*}
Combining this with (4.12) and (4.14), we obtain
\begin{equation*}
\left\| u\right\| _{H^{1}\left( Q_{\delta }\right) }\leq \frac{Ce^{4\lambda
m}}{\gamma }K^{2}+C\left\| u\right\| _{H^{1}\left( Q_{T}\right)
}^{2}e^{-2\lambda \delta }
\end{equation*}
\begin{equation*}
+\chi _{\varphi }C\gamma \left\| u\left( x,0\right) \right\| _{H^{1}\left(
\Omega \right) }^{2}+\chi _{\psi }C\gamma \left\| u_{t}\left( x,0\right)
\right\| _{L_{2}\left( \Omega \right) }^{2}.
\end{equation*}
Hence, there exists a number $t^{\ast }\in \left( 0,\delta \right) $ such
that
\begin{equation*}
\int\limits_{\Omega }\left( \left| \nabla _{x,t}u\right| ^{2}+u^{2}\right)
\left( x,t^{\ast }\right) dx\leq \frac{Ce^{4\lambda m}}{\delta \gamma }K^{2}+%
\frac{C}{\delta }\left\| u\right\| _{H^{1}\left( Q_{T}\right)
}^{2}e^{-2\lambda \delta }
\end{equation*}

\begin{equation*}
+\frac{1}{\delta }\left[ \chi _{\varphi }C\gamma \left\| u\left( x,0\right)
\right\| _{H^{1}\left( \Omega \right) }^{2}+\chi _{\psi }C\gamma \left\|
u_{t}\left( x,0\right) \right\| _{L_{2}\left( \Omega \right) }^{2}\right] .
\end{equation*}
This inequality combined with (4.1) and the standard energy estimates
implies that
\begin{equation*}
\left\| u\right\| _{H^{1}\left( Q_{T}\right) }^{2}+\chi _{\varphi }\left\|
u\left( x,0\right) \right\| _{H^{1}\left( \Omega \right) }^{2}+\chi _{\psi
}\left\| u_{t}\left( x,0\right) \right\| _{L_{2}\left( \Omega \right)
}^{2}\leq C\left\| u\right\| _{H^{1}\left( Q_{T}\right) }^{2}e^{-2\lambda
\delta }
\end{equation*}
\begin{equation}
+\frac{Ce^{4\lambda m}}{\gamma }K^{2}+\chi _{\varphi }C\gamma \left\|
u\left( x,0\right) \right\| _{H^{1}\left( \Omega \right) }^{2}+\chi _{\psi
}C\gamma \left\| u_{t}\left( x,0\right) \right\| _{L_{2}\left( \Omega
\right) }^{2}.  \tag{4.15}
\end{equation}
Note that $\delta $ is independent on $\lambda .$ Choose sufficiently large $%
\lambda _{0}$ such that
\begin{equation*}
1-Ce^{-2\lambda _{0}\delta }>\frac{1}{2}
\end{equation*}
and set $\lambda :=\lambda _{0}.$ Choose $\gamma $ so small that $C\gamma
<1/2.$ Then we obtain (4.2) from (4.15). $\square $

\section{Convergence}

Theorem 4.1 enables us to prove convergence of our method. Following the
Tikhonov concept for ill-posed problems [18], we first introduce an
``ideal'' exact solution of either $\varphi $ or $\psi $ problem without an
error in the data. Next, we assume the existence of the error in \ the
boundary data $f$ and $g$ and prove that our solution tends to the exact one
as the level of error in the data tends to zero. We consider the more
general Inverse Problem 1. Let $f^{\ast }\in H^{1}\left( S_{T}\right) $ and $%
g^{\ast }\in L_{2}\left( S_{T}\right) $ be the exact boundary data (1.3), $%
F^{\ast }\in L_{2}\left( Q_{T}\right) $ be the exact right hand side of
equation (1.1) and $\varphi ^{\ast }$ and $\psi ^{\ast }$ be exact initial
conditions. We assume that there exists an exact function $u^{\ast }\in
H^{2}\left( Q_{T}\right) $ satisfying
\begin{equation}
u_{tt}^{\ast }=L(x,t)u^{\ast }+F^{\ast }\left( x,t\right) \text{ in }Q_{T},
\tag{5.1}
\end{equation}
with initial conditions
\begin{equation}
u^{\ast }\left( x,0\right) =\varphi ^{\ast }\left( x\right) ,u_{t}^{\ast
}\left( x,0\right) =\psi ^{\ast }\left( x\right) ,\varphi ^{\ast }\in
H^{1}\left( \Omega \right) ,\psi ^{\ast }\in L_{2}\left( \Omega \right) ,
\tag{5.2}
\end{equation}
\begin{equation}
u^{\ast }\mid _{S_{T}}=f^{\ast }\left( x,t\right) ,\frac{\partial u^{\ast }}{%
\partial \nu }\mid _{S_{T}}=g^{\ast }\left( x,t\right) ,  \tag{5.3}
\end{equation}
where $\varphi ^{\ast }$ and $\psi ^{\ast }$ are exact initial conditions.
We assume that the real boundary data in (1.3) have an error, so as the
given initial condition. In other words, we assume that
\begin{equation}
\left\| f-f^{\ast }\right\| _{H^{1}\left( S_{T}\right) }+\left\| g-g^{\ast
}\right\| _{L_{2}\left( S_{T}\right) }+\left\| F-F^{\ast }\right\|
_{L_{2}\left( Q_{T}\right) }  \tag{5.4}
\end{equation}
\begin{equation*}
+\chi _{\psi }\left\| \varphi -\varphi ^{\ast }\right\| _{H^{1}\left( \Omega
\right) }+\chi _{\varphi }\left\| \psi -\psi ^{\ast }\right\| _{L_{2}\left(
\Omega \right) }\leq \delta ,
\end{equation*}
where $\delta >0$ is a small number. The following convergence theorem holds

\textbf{Theorem 5.1.} \emph{Suppose that }$T>R.$\emph{\ Let }$u_{\varepsilon
\delta }\in H^{2}\left( Q_{T}\right) $\emph{\ be the solution of the QRM
problem (3.2), which is guaranteed by Lemma 2.1. Let conditions (5.1)-(5.4)
be satisfied. Then the following estimate is valid}
\begin{equation*}
\left\| u-u^{\ast }\right\| _{H^{1}\left( Q_{T}\right) }+\chi _{\varphi
}\left\| \varphi -\varphi ^{\ast }\right\| _{H^{1}\left( \Omega \right)
}+\chi _{\psi }\left\| \psi -\psi ^{\ast }\right\| _{L_{2}\left( \Omega
\right) }\leq C\left( \delta +\sqrt{\varepsilon }\right) .
\end{equation*}

\textbf{Proof. }Since the functional $J_{0}\left( u\right) $ with the exact
data (5.2), (5.3) achieves its minimal zero value at $u:=u^{\ast },$ then
the function $u^{\ast }$ satisfies equation (3.2) with $\varepsilon =0$ and
with the exact data (5.2), (5.3). Subtracting that equation for $u^{\ast }$
from equation \ (3.2) for $u:=u_{\varepsilon \delta },$ denoting $%
w=u_{\varepsilon \delta }-u^{\ast }$, setting in resulting equation $v:=w$
and using (5.4), we obtain similarly with (3.3)
\begin{equation*}
\int\limits_{Q_{T}}\left( Mw\right) ^{2}dxdt+\chi _{\psi }\left\|
w(x,0)\right\| _{H^{1}\left( \Omega \right) }^{2}+\chi _{\varphi }\left\|
w_{t}(x,0)\right\| _{L_{2}\left( \Omega \right) }^{2}
\end{equation*}
\begin{equation*}
+\left\| w\mid _{S_{T}}\right\| _{H^{1}\left( S_{T}\right) }^{2}+\left\|
w_{\nu }\mid _{S_{T}}\right\| _{L_{2}\left( S_{T}\right) }^{2}\leq 4\delta
^{2}+\varepsilon .
\end{equation*}
The \ rest of the proof follows immediately from Theorem 4.1. $\square $

\section{Numerical Implementation}

In our numerical study we have considered the Inverse Problem 2. To generate
the data for the inverse problem, we have solved the Cauchy problem
\begin{equation}
u_{tt}=\Delta u,\left( x,t\right) \in \mathbb{R}^{2}\times \left( 0,T\right)
,  \tag{6.1}
\end{equation}
\begin{equation}
u(x,0)=\varphi \left( x\right) ,u_{t}(x,0)=\psi \left( x\right) .  \tag{6.2}
\end{equation}
In our numerical experiments $\psi \left( x\right) \equiv 0$ for the $%
\varphi -$problem, and $\varphi \left( x\right) \equiv 0$ for the $\psi -$%
problem. Because of (1.5) and the finite speed of propagation, we use in our
solution of the forward problem zero Dirichlet boundary condition at the
boundary of the rectangle $\left( -T,a+T\right) \times \left( -T,a+T\right) $
(Figure 1). Hence, we solve initial boundary value problem inside of this
rectangle for equation (6.1) with initial conditions (6.2) at zero Dirichlet
boundary condition. In all our calculations we took $a=1.$ In tests 1, 2 and
5, which are concerned with the Inverse Problem 2, we took $T=3.$ Hence,
condition (1.9) is satisfied. Tests 3 and 4 are concerned with the Inverse
Problem 1 and we have taken different values of $T$ in these tests. The
square $SQ(a)$ is $SQ(a)=SQ(1)=\left( 0,1\right) \times \left( 0,1\right) ,$
the domain $\Omega $ in tests 1,2 and 5 is
\begin{equation}
\Omega :=\left( 0,4\right) \times \left( 0,4\right)  \tag{6.3}
\end{equation}
and in all tests
\begin{equation}
\varphi \left( x\right) =\psi \left( x\right) =0\text{ for }x\notin SQ(1).
\tag{6.4}
\end{equation}

We have solved the Cauchy problem (6.1), (6.2) via finite differences using
the uniform grid. We set
\begin{equation*}
u(t_{k},x_{1n},x_{2m})\approx
u_{kmn},\,k=0,...,N_{t},\,\,n=0,...,N_{x},m=0,...,N_{y},
\end{equation*}
\begin{equation*}
t_{k}=kh_{t},\,x_{1n}=nh_{x_{1}},\,x_{2m}=mh_{x_{2}},
\end{equation*}
step sizes $h_{x_{1}}=h_{x_{2}}=0.1,h_{t}=1/15$ and $N_{x}=N_{y}=10,N_{t}=45$%
. This solution has generated the boundary data (1.8). Next, we have
introduced noise in these data as
\begin{equation}
f_{n}\left( x^{i},t_{j}\right) =f\left( x^{i},t_{j}\right) \left( 1+\gamma
N\left( t_{j}\right) \right) ,g_{n}\left( x^{i},t_{j}\right) =g\left(
x^{i},t_{j}\right) \left( 1+\gamma N\left( t_{j}\right) \right) ,  \tag{6.5}
\end{equation}
where $\left( x^{i},t_{j}\right) $ is the grid point at the boundary. Here $%
N\in \left( -1,1\right) $ is a pseudo random variable, which is given by
function $Math.random()$ in Java and $\gamma \in \left[ 0.05,0.5\right] $ is
the noise level. We have chosen the grid points the same as ones in the
finite difference scheme we have solved the problem (6.1), (6.2). The
presence of the random noise in the date prevents us from committing
``inverse crime''. In (6.5) points $x^{i}\in \Gamma _{1T}\cup \Gamma _{2T}.$
As to $\Gamma _{3T}\cup \Gamma _{4T},$ we simply set $f=g=0$ on this part of
the boundary, because of (1.7).

To find the minimizer of the functional $J_{\varepsilon },$ we have also
used finite differences. We have used in (3.1) the finite difference
approximations for $Mu=u_{tt}-\Delta u$ and $u_{\nu }\mid _{S_{T}}$ and have
minimized the resulting functional $\widetilde{J}_{\varepsilon }$ with
respect to the vector $\left\{ u_{kmn}\right\} ,$ which approximates values
of the function $u$ at grid points. Here $\widetilde{J}_{\varepsilon }$
means the functional $J_{\varepsilon },$ which\ is expressed via the finite
differences. The norms $\left\| u_{x_{1}}(x,0)\right\| _{L_{2}\left( \Omega
\right) },$ $\left\| u_{x_{2}}(x,0)\right\| _{L_{2}\left( \Omega \right) }$
in $\left\| u(x,0)\right\| _{H^{1}\left( \Omega \right) }$ in the $\psi -$%
problem were calculated via finite differences. As to the term $\left\|
D^{\beta }u\mid _{S_{T}}-D^{\beta }f\right\| _{L_{2}\left( S_{T}\right)
}^{2} $ in (3.1), we have used only $\beta =0,$ thus ending up with $\left\|
u\mid _{S_{T}}-f\right\| _{L_{2}\left( S_{T}\right) }^{2}$ (in the discrete
sense)$.$ Note that since $\beta =0,$ our numerical results seem to be
stronger than Theorem 4.1 predicts. The integrals were calculated as
\begin{equation*}
\int\limits_{\Omega _{T}}(u_{tt}-\Delta u)^{2}dv\approx \frac{%
h_{t}h_{x_{2}}h_{x_{1}}}{h_{t}^{4}}\sum_{k=1}^{N_{t}-1}\sum_{m=1}^{N_{y}-1}%
\sum_{n=1}^{N_{x}-1}M_{kmn}^{2},
\end{equation*}
where
\begin{multline*}
M_{kmn}=(u_{k+1,mn}-2u_{kmn}+u_{k-1,mn})-\lambda
_{y}(u_{k,m+1,n}-2u_{kmn}+u_{k,m-1,n}) \\
-\lambda _{x}\left( u_{km,n+1}-2u_{kmn}+u_{km,n-1}\right) \\
=(u_{k+1,mn}+u_{k-1,mn})-\lambda _{y}(u_{k,m+1,n}+u_{k,m-1,n})-\lambda
_{x}(u_{km,n+1}+u_{km,n-1})-\lambda _{t}u_{kmn},
\end{multline*}
where
\begin{equation*}
\lambda _{x}=\frac{h_{t}^{2}}{h_{x_{1}}^{2}},\quad \lambda _{y}=\frac{%
h_{t}^{2}}{h_{x_{2}}^{2}},\quad \lambda _{t}=2(1-\lambda _{x}-\lambda _{y}).
\end{equation*}
Also,
\begin{equation*}
\int\limits_{0}^{T}\int\limits_{0}^{a+T}(u(t,x_{2},x_{1}^{\ast
})-f(t,x_{2}))^{2}dx_{2}dt\approx
h_{t}h_{x_{2}}\sum_{k=0}^{N_{t}}\sum_{m=0}^{N_{x_{2}}}H_{km}^{2},
\end{equation*}
where
\begin{equation*}
H_{km}=u_{kmn_{\ast }}-h_{km},
\end{equation*}
where $n_{\ast }$ in the layer number (value of $x_{1}^{\ast }$) at which
the grid function $f_{km}$ is given.

To minimize the functional $\widetilde{J}_{\varepsilon },$ we have used the
conjugate gradient method. Derivatives with respect to variables $u_{kmn}$
where calculated in closed forms, using the following formula
\begin{equation*}
\frac{\partial u_{kmn}}{\partial u_{\overline{k}\overline{m}\overline{n}}}%
=\delta _{k\overline{k}}\delta _{m\overline{m}}\delta _{n\overline{n}},
\end{equation*}
where $\delta _{k\overline{k}}$ is the Kronecker symbol. This formula can be
conveniently used to obtain closed form expressions for derivatives
\begin{equation*}
\frac{\partial \widetilde{J}_{\varepsilon }\left( u\right) }{\partial u_{kmn}%
}.
\end{equation*}

\begin{figure}[t]
\begin{center}
\begin{tabular}{c@{\hspace{0.05cm}}c}
\includegraphics[width=8.5cm]{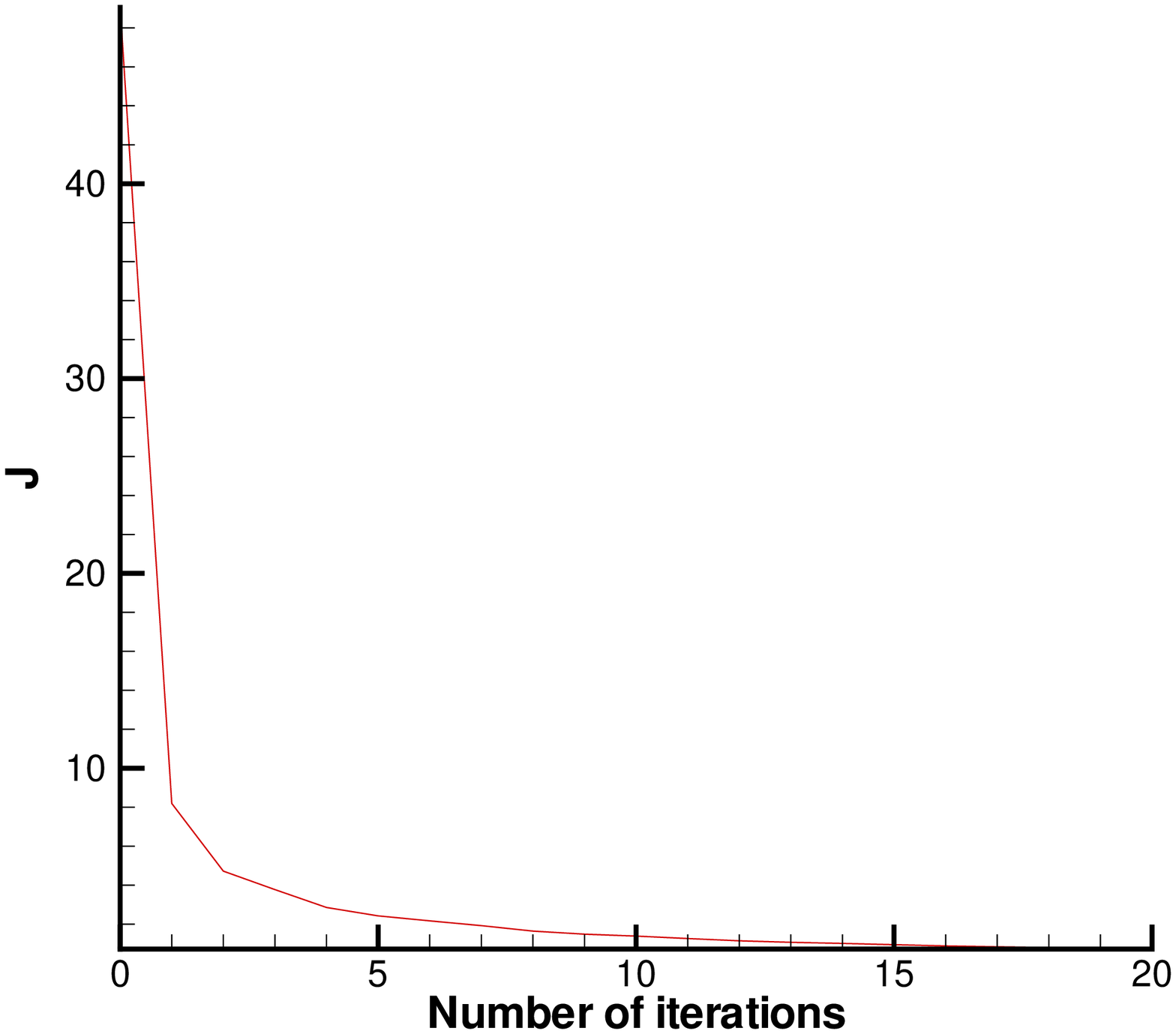} & \includegraphics[width=8.5cm]{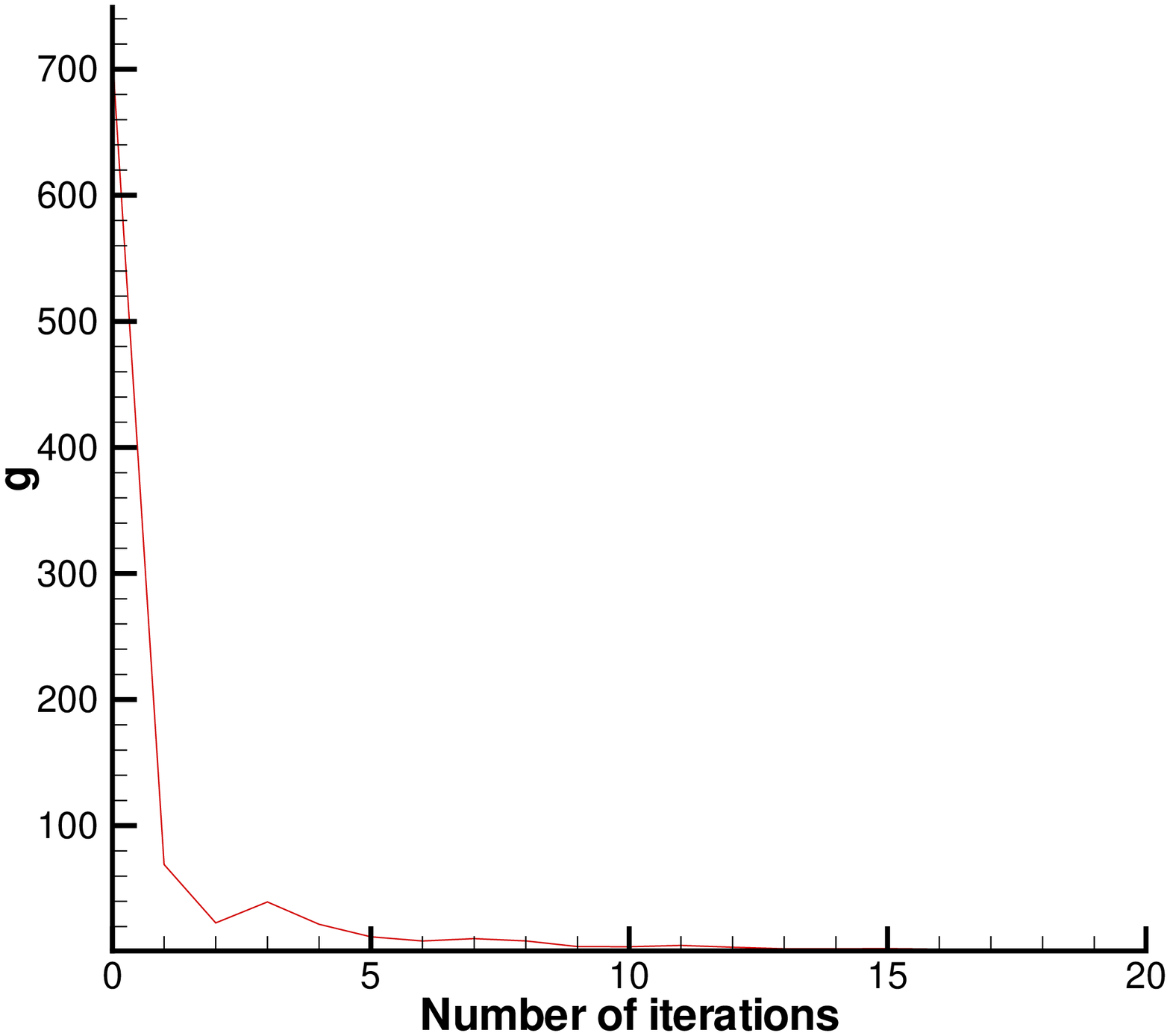}
\end{tabular}
\end{center}
\caption{Typical dependence of the functional $J=J_{\protect\varepsilon }$
(left) and $g=\|\protect\nabla J_{\protect\varepsilon }\|^2$ (right) on
number of iterations.}
\label{fig:functl_and_grad}
\end{figure}

Let $a$ be the vector of unknowns of the functional $\widetilde{J}%
_{\varepsilon }$. We start our iterative process from $a:=a_{0}=0$. It is
well known in the field of ill-posed problems that the number of iterations
can often be taken as a regularization parameter, and it depends, of course
on the range of parameters of a problem one considers. We have found that
the optimal number of iterations for our range of parameters is $300$. Thus,
in all numerical examples below 300 iterations of the conjugate gradient
method were used, thus ending up with $a_{300}$. Figure $\ref
{fig:functl_and_grad}$ displays typical dependencies of the functional $%
\widetilde{J}_{\varepsilon }\left( a_{k}\right) $ and the norm of its
gradient on the iteration number $k$.

\section{Numerical Results}

In this section we present results of some numerical experiments. We have
always used $\varepsilon =10^{-6}.$ Larger values of $\varepsilon $ such as $%
10^{-5}$ brought lower quality results. In our numerical experiments we have
imaged both smooth slowly varying functions and the finite difference
analogue of the $\delta -$ function. Let $\left( x_{1k},x_{2r}\right) \in
\Omega $ be a fixed grid point. To obtain the finite difference analogue of $%
\delta \left( x_{1}-x_{1k},x_{2}-x_{2r}\right) $, we consider the following
grid points $\left( x_{1n},x_{2m}\right) $ and we model the function $\delta
\left( x_{1n}-x_{1k},x_{2m}-x_{2r}\right) $ as
\begin{equation*}
\delta \left( x_{1n}-x_{1k},x_{2m}-x_{2r}\right) =\frac{3}{%
4h_{x_{2}}h_{x_{1}}}\delta _{nk}\delta _{mr},
\end{equation*}
where the multiplier at $\delta _{nk}\delta _{mr}$ is chosen such that the
volume of the pyramid based on $\left( x_{1k-1},x_{2r-1}\right) ,\,\left(
x_{1k-1},x_{2r+1}\right) ,\,\left( x_{1k+1},x_{2r+1}\right) ,\,\left(
x_{1k+1},x_{2r-1}\right) $ equals to $1$. Hence, the support of the function
$\delta \left( x_{1n}-x_{1k},x_{2m}-x_{2r}\right) $ is limited only to the
point $\left( x_{1n},x_{2m}\right) $.
\begin{figure}[tbp]
\begin{center}
\includegraphics[width=16cm, height=8cm]{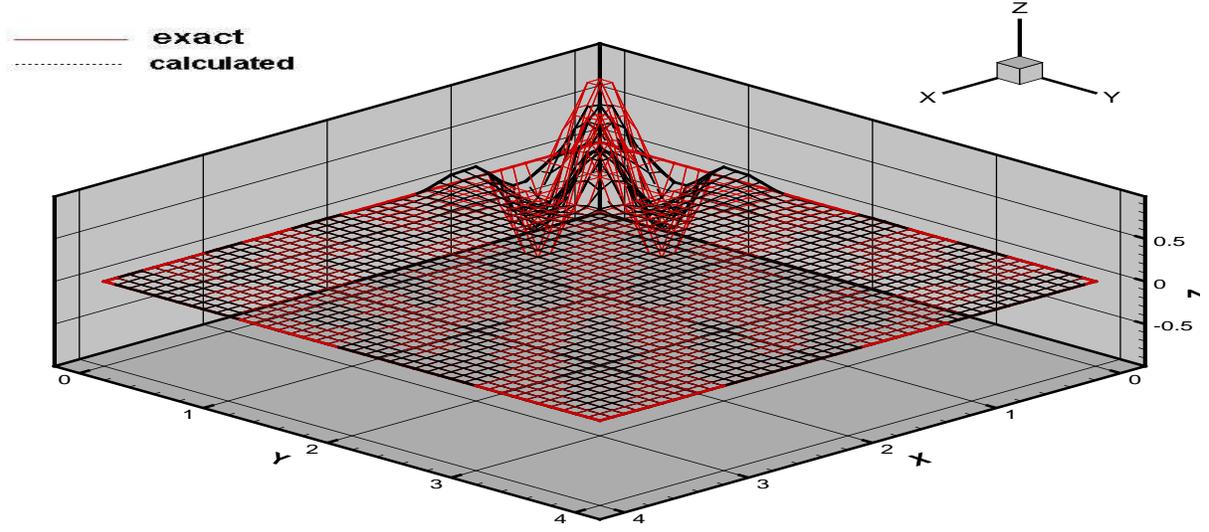}
\end{center}
\caption{Exact (red) and calculated (black) functions $\protect\varphi $ in
(7.1) without balancing coefficients with 5\% noise in boundary data.}
\label{fig:bad_bndry}
\end{figure}
\begin{figure}[tbp]
\begin{center}
\includegraphics[width=16cm, height=8cm]{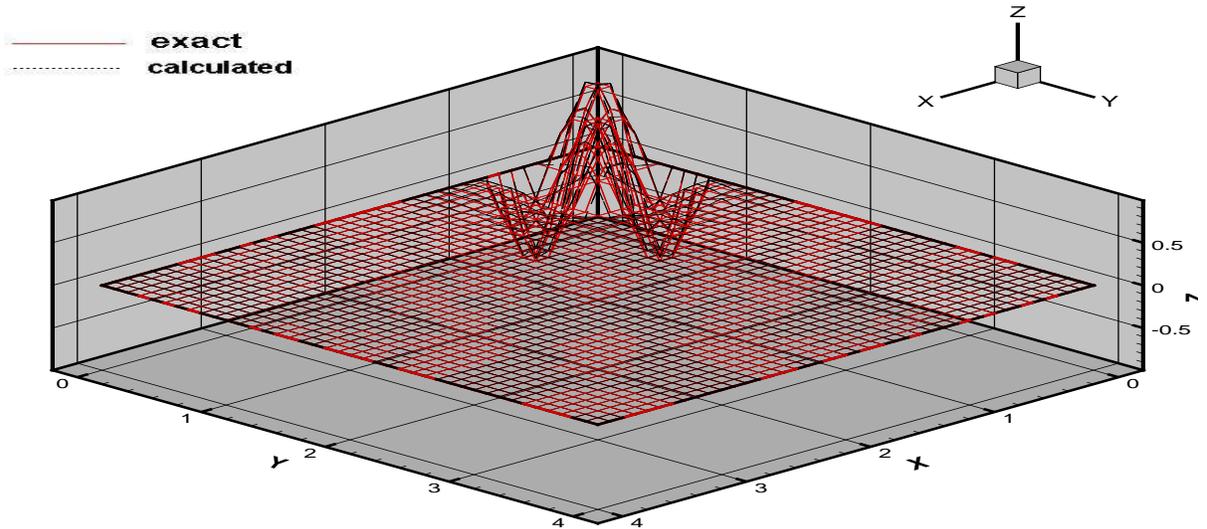}
\end{center}
\caption{Exact (red) and calculated (black) functions $\protect\varphi $ in
(7.1) with balancing coefficients with 5\% noise in boundary data.}
\label{fig:good_bndry}
\end{figure}

We have observed that having equal coefficient at all terms of the
functional $\widetilde{J}_{\varepsilon }$ in (3.1) does not lead to good
reconstruction results. This is because not all the terms of (3.1) provide
an equal impact in this functional. For example, for the $\varphi -$problem
with no noise in the data for the function
\begin{equation}
\varphi \left( x\right) =\left\{
\begin{array}{c}
\sin (2\pi x_{1})\sin (2\pi x_{2}),x\in SQ\left( 1\right) \\
0\text{ otherwise}
\end{array}
\right\}  \tag{7.1}
\end{equation}
we first got the result displayed in Figure $\ref{fig:bad_bndry}$. One can
observe that the error at the boundary is significant. And indeed, the
values of two terms in (3.1) after 300 iterations were for this case
\begin{equation*}
\int\limits_{\Omega _{T}}(u_{tt}-\Delta u)^{2}dxdt\approx 10^{-3},\quad
\Vert u-f\Vert _{L_{2}(S_{T})}\approx 10^{-2}.
\end{equation*}
Hence, the impact of the boundary term in (3.1) is 10 greater than the
impact of the $\left\| Mu\right\| _{L_{2}\left( Q_{T}\right) }^{2}.$ To
minimize the error at the boundary, we took the balancing coefficient $1000$
at $1000\cdot \Vert u-f\Vert _{L_{2}(S_{T})}$ instead of $1\cdot \Vert
u-f\Vert _{L_{2}(S_{T})}$. The other balancing coefficients equal to $1$.
The quality of the resulting image was improved, see Figure $\ref
{fig:good_bndry}$. Thus, in all our tests with the $\varphi -$problem we
have taken the same balancing coefficients. In the case of the $\psi -$%
problem we have taken $100\cdot \chi _{\psi }\left\| u(x,0)-\varphi \right\|
_{H^{1}\left( \Omega \right) }^{2}$ and the other balancing coefficients
equal to $1$.

Note that Theorems 4.1 and 5.1 remain the same, including their proofs, if
balancing coefficients are introduced.

\begin{figure}[tbp]
\begin{center}
\includegraphics[width=16cm, height=8cm]{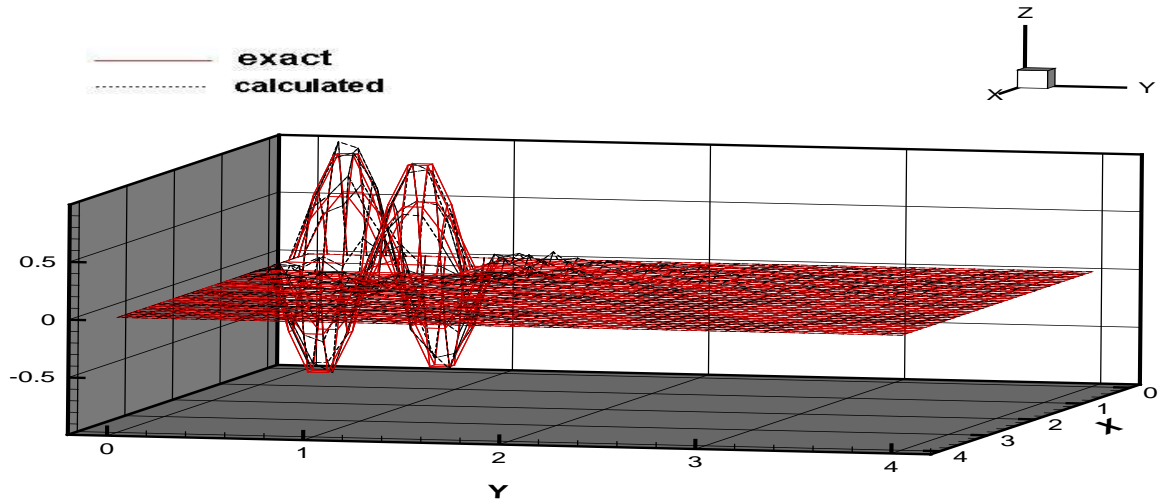}
\end{center}
\caption{Test 1. Exact (red) and calculated (black) functions $\protect%
\varphi $ in (7.1) with 25\% noise in the boundary data.}
\label{fig:phi_sin_25}
\end{figure}

\begin{figure}[tbp]
\begin{center}
\includegraphics[width=16cm, height=8cm]{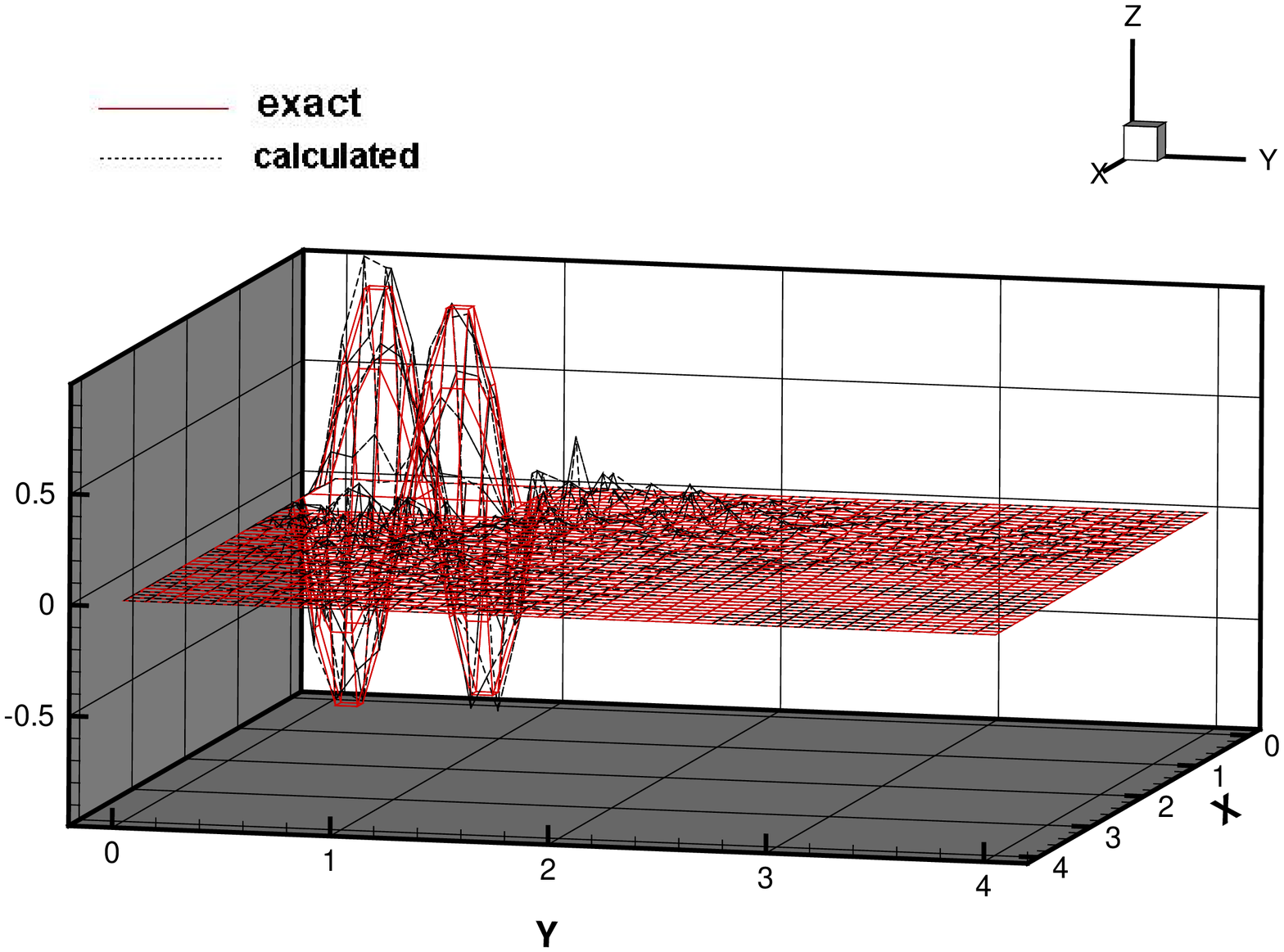}
\end{center}
\caption{Test 1. Exact (red) and calculated (black) functions $\protect%
\varphi $ in (7.1) with 50\% noise in the boundary data.}
\label{fig:phi_sin_50}
\end{figure}

\begin{figure}[tbp]
\begin{center}
\includegraphics[width=16cm, height=16 cm]{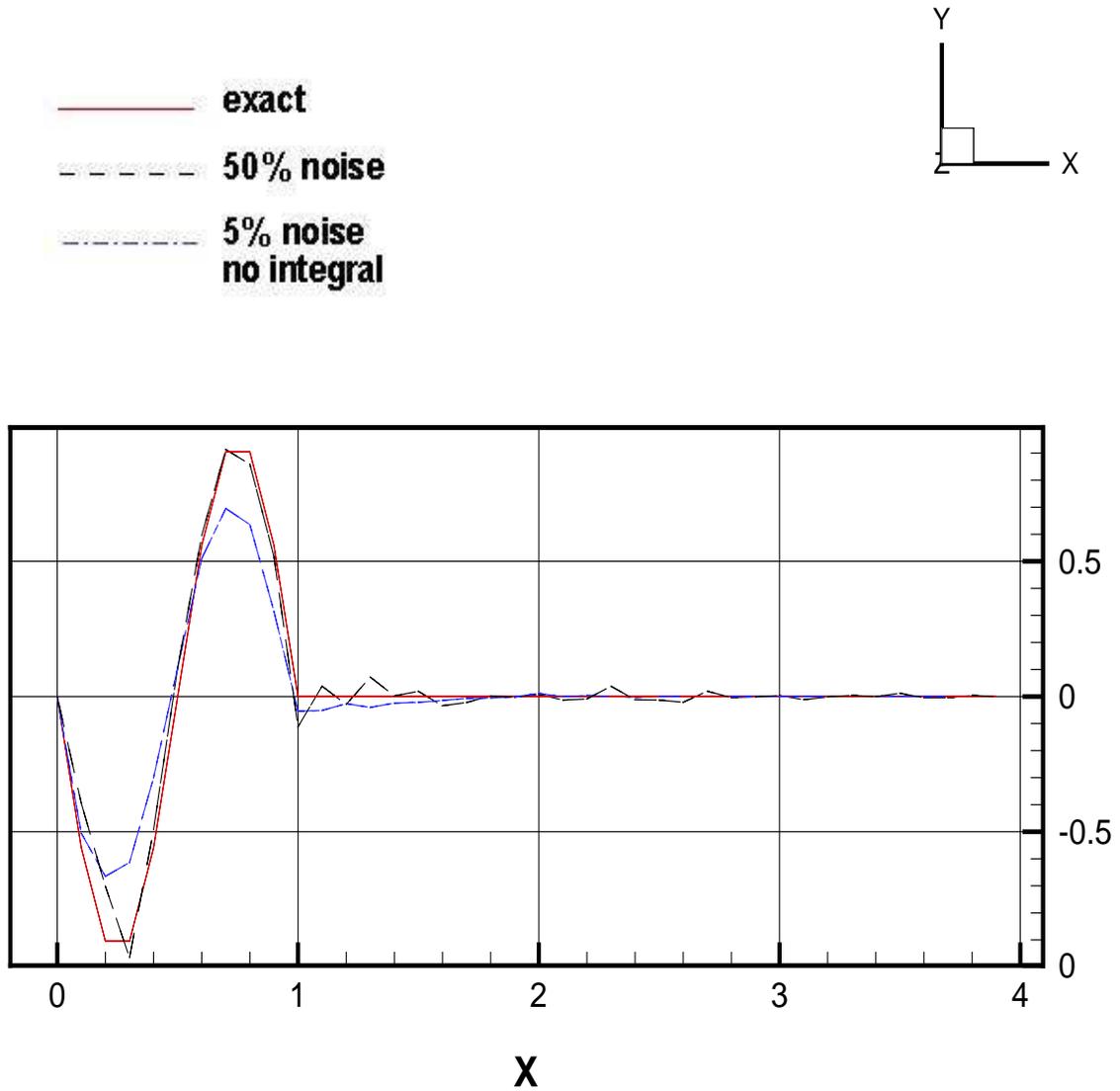}
\end{center}
\caption{Test 1. Cross sections of exact (red) and calculated (black, blue)
functions $\protect\varphi $ with 5\%, 50\% noise, ''no integral'' means $%
\protect\chi _{\protect\varphi =0}$. One can see that the maximal value of
the case $\protect\chi _{\protect\varphi =0}$ is $0.7/(-0.7)$. The maximal
value of the exact function is $0.9<1$ only because of the grid step size.}
\label{fig:phi_sin_cross}
\end{figure}

\textbf{Test 1.} \emph{The }$\varphi -$\emph{problem.} Here $\psi \left(
x\right) \equiv 0$ and the function $\varphi \left( x\right) $ to be
reconstructed is one in (7.1). In Figures $\ref{fig:phi_sin_25}$ and $\ref
{fig:phi_sin_50}$ represent resulting images with 25\% and 50\% noise
respectively. Next, we test our method for the case when the term with $\chi
_{\varphi }$ is absent in the functional $J_{\varepsilon }\left( u\right) $
in (3.1).\ Regardless on the small amount of noise in the data, both maximal
($1$) and minimal ($-1$) values of the imaged function were missed by about
22\% in this case, whereas they were not missed in the previous cases with
25\% and 50\% noise when the term with $\chi _{\varphi }$ was not absent in
(3.1). To see this, we display on Figure $\ref{fig:phi_sin_cross}$ the
1-dimensional cross-sections by the straight line $\left\{ x_{1}=0.5\right\}
$ of the correct function (7.1), the imaged function with 50\% noise of
Figure $\ref{fig:phi_sin_50}$ and the imaged function with the absent term
with $\chi _{\varphi }$ and 5\% noise. One can observe that the maximal
value of the calculated function is $0.7$, while the maximal absolute value
of the correct function is $0.9$, so as the one of Figure $\ref
{fig:phi_sin_50}$. Here we have $0.9$ instead of $1$ only because the points
with the absolute value of $1$ are not the grid points. \ This emphasizes
the importance of the incorporation of the term with $\chi _{\varphi }.$ We
have observed the same for the $\psi -$ problem (images not shown).

\begin{figure}[tbp]
\begin{center}
\includegraphics[width=16cm, height=8cm]{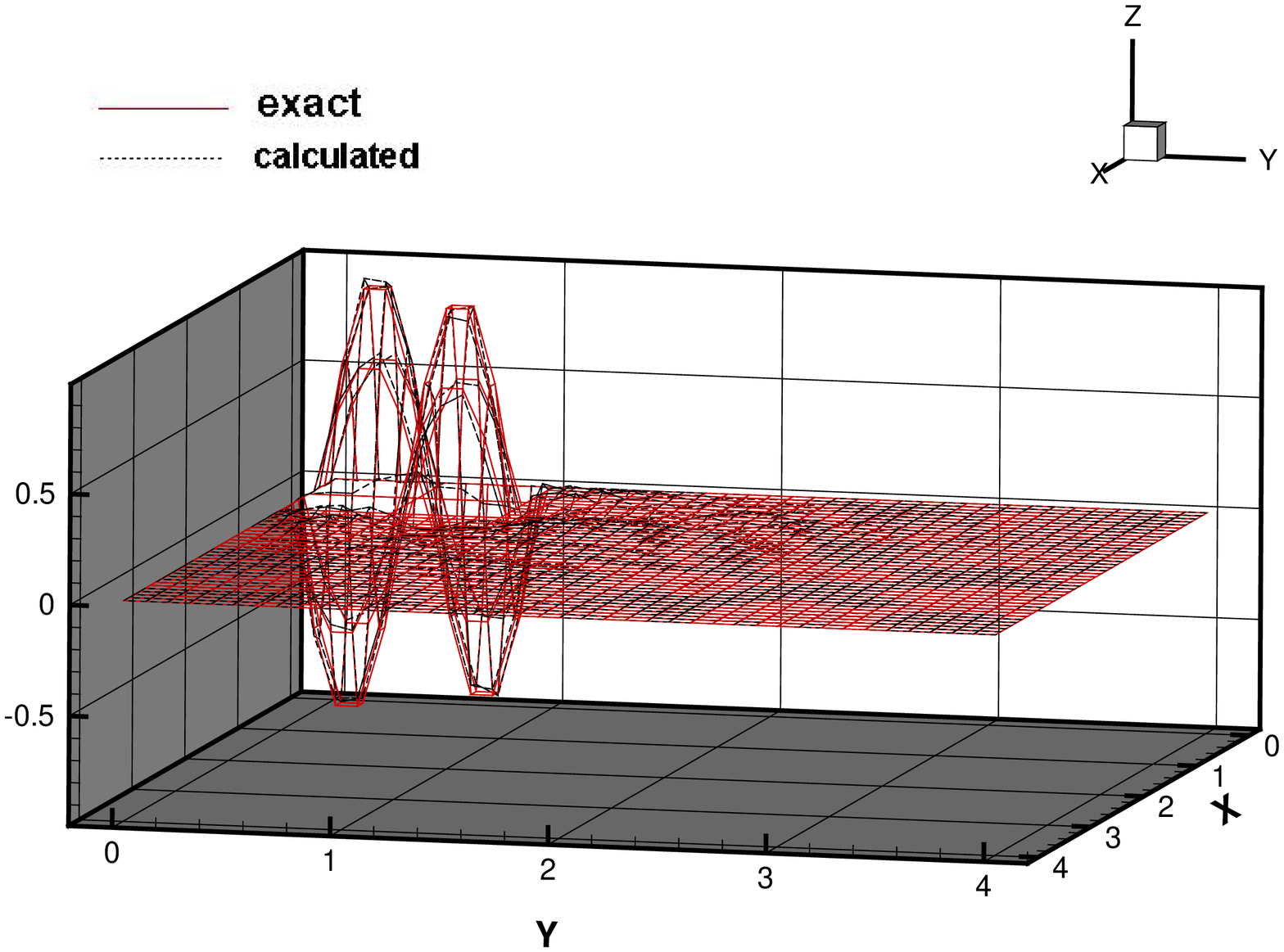}
\end{center}
\caption{Test 2. Exact (red) and calculated (black) functions $\protect\psi$
in (7.2) with 5\% noise in the boundary data.}
\label{fig:psi_sin_5}
\end{figure}

\begin{figure}[tbp]
\begin{center}
\includegraphics[width=16cm, height=8cm]{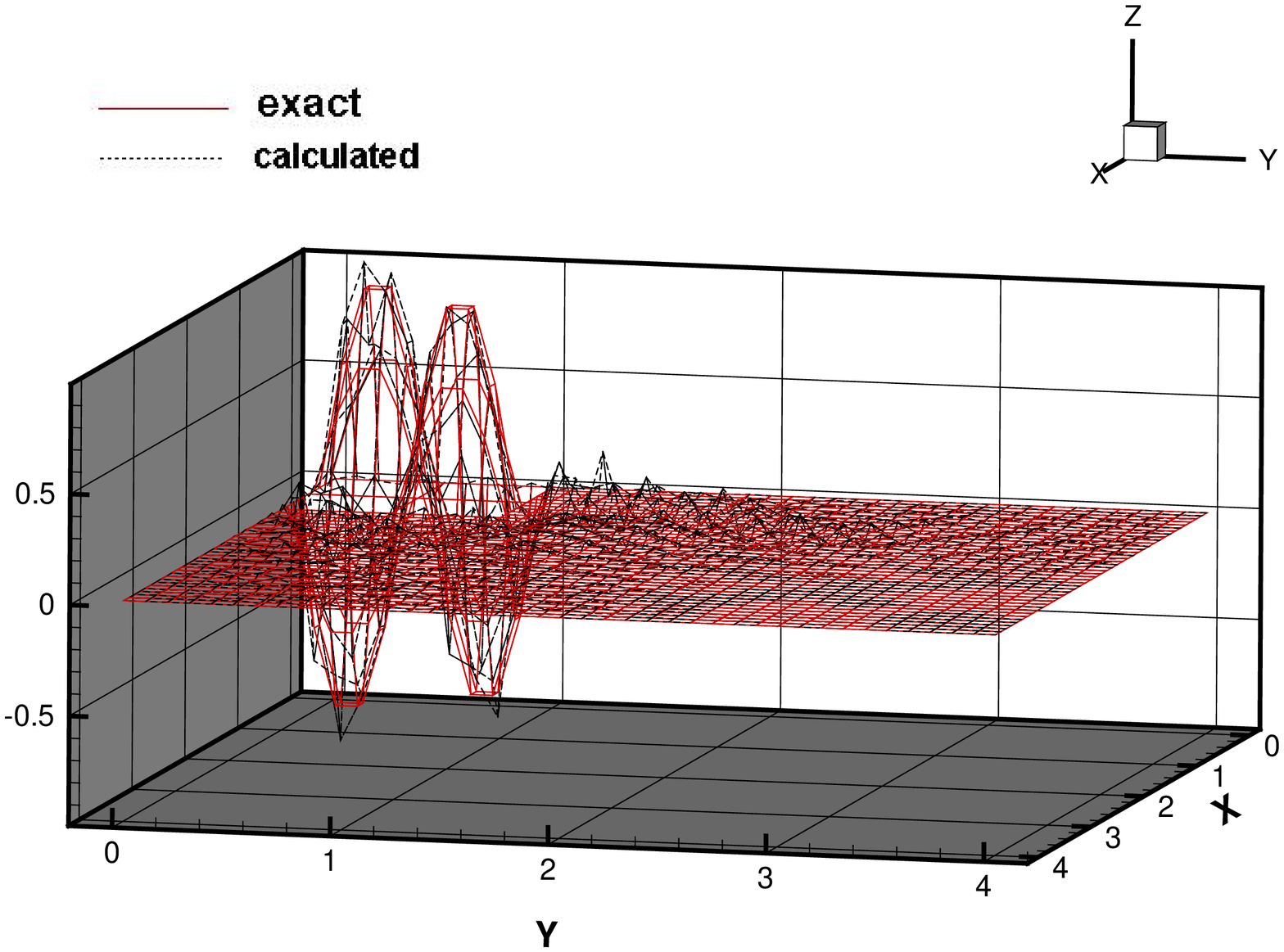}
\end{center}
\caption{Test 2. Exact (red) and calculated (black) functions $\protect\psi$
in (7.2) with 25\% noise in the boundary data.}
\label{fig:psi_sin_25}
\end{figure}

\begin{figure}[tbp]
\begin{center}
\includegraphics[width=16cm, height=8cm]{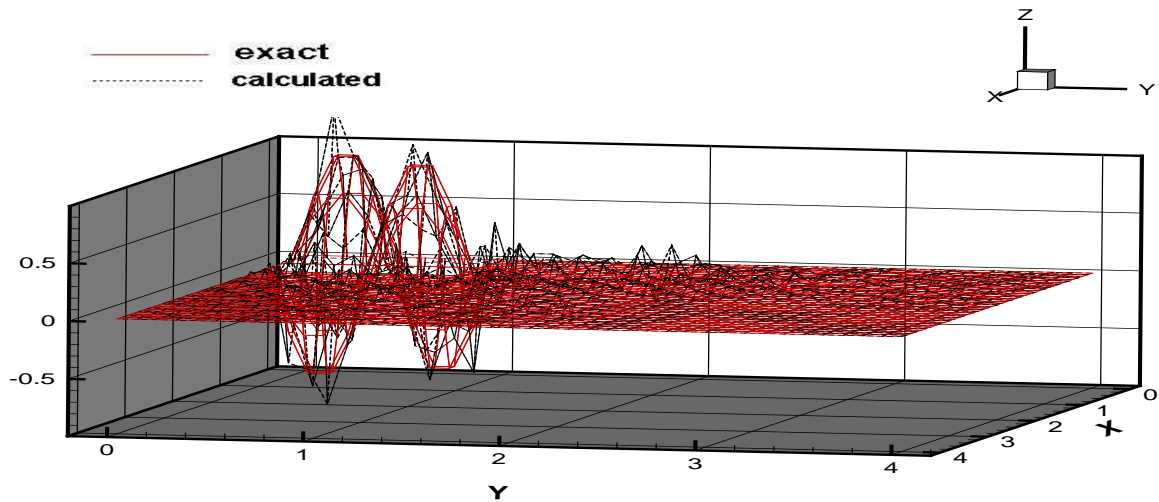}
\end{center}
\caption{Test 2. Exact (red) and calculated (black) functions $\protect\psi$
with 50\% noise in the boundary data.}
\label{fig:psi_sin_50}
\end{figure}

\textbf{Test 2}.\emph{\ The }$\psi -$\emph{problem.} In this case $\varphi
\left( x\right) \equiv 0$ and the function $\psi \left( x\right) $ to be
reconstructed is
\begin{equation}
\psi \left( x\right) =\left\{
\begin{array}{c}
\sin (\frac{\pi }{2}\left( x_{1}-0.5\right) \sin (\frac{\pi }{2}\left(
x_{2}-0.5\right) ),x\in SQ\left( 1\right) \\
0\text{ otherwise.}
\end{array}
\right\} .  \tag{7.2}
\end{equation}
Figures $\ref{fig:psi_sin_5}$, $\ref{fig:psi_sin_25}$ and $\ref
{fig:psi_sin_50}$display resulting images of the function (7.2) with 5\%,
25\% and 50\% of the noise level in the data respectively.

\begin{figure}[tbp]
\begin{center}
\includegraphics[width=16cm]{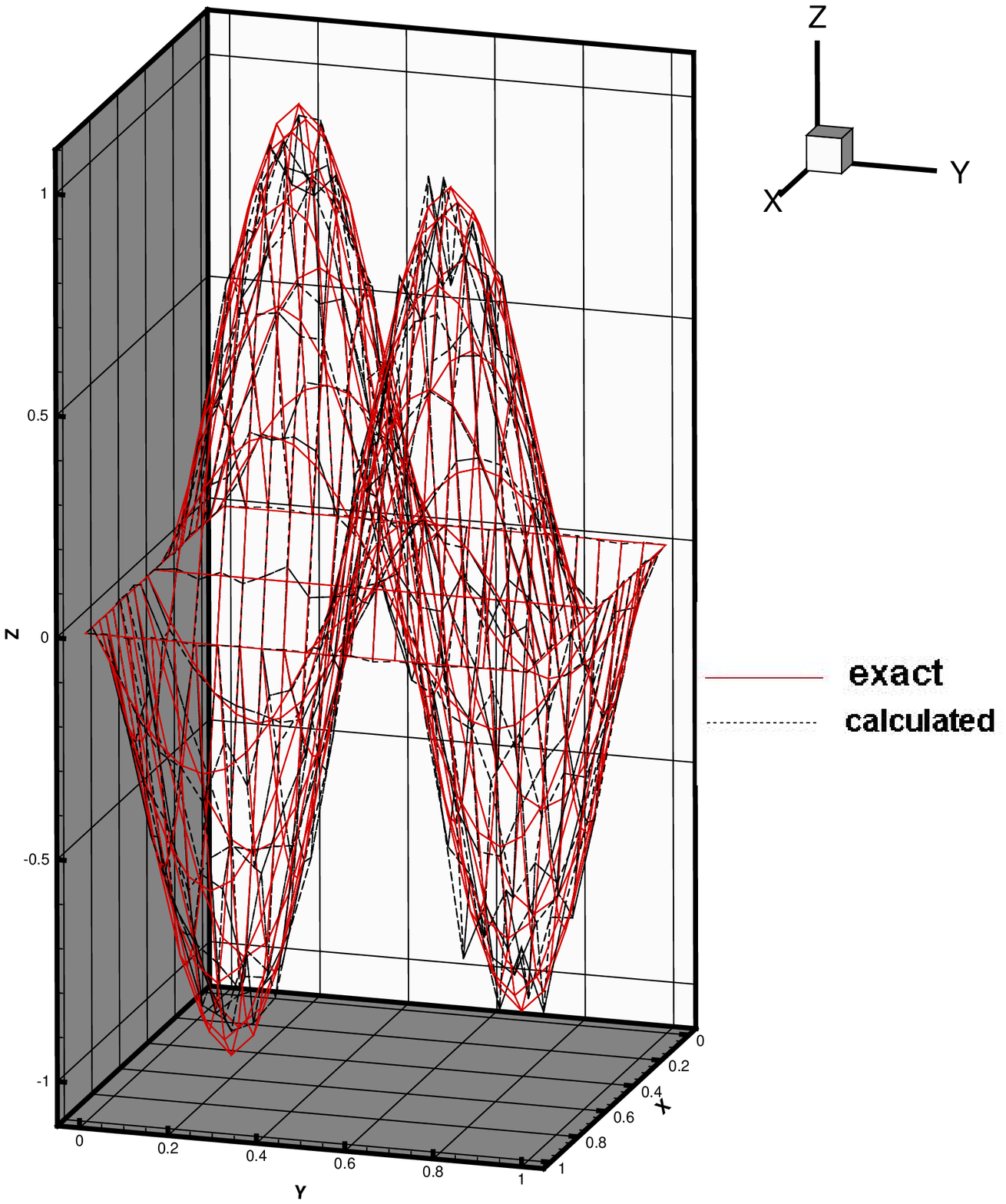}
\end{center}
\caption{Test 3. Exact (red) and calculated (black) solutions of the problem
$\protect\varphi -$ in SQ(1) with 25\% noise in the boundary data for $%
T=0.75 $.}
\label{fig:phi_sin_e_t_25}
\end{figure}

\begin{figure}[tbp]
\begin{center}
\includegraphics[width=16cm]{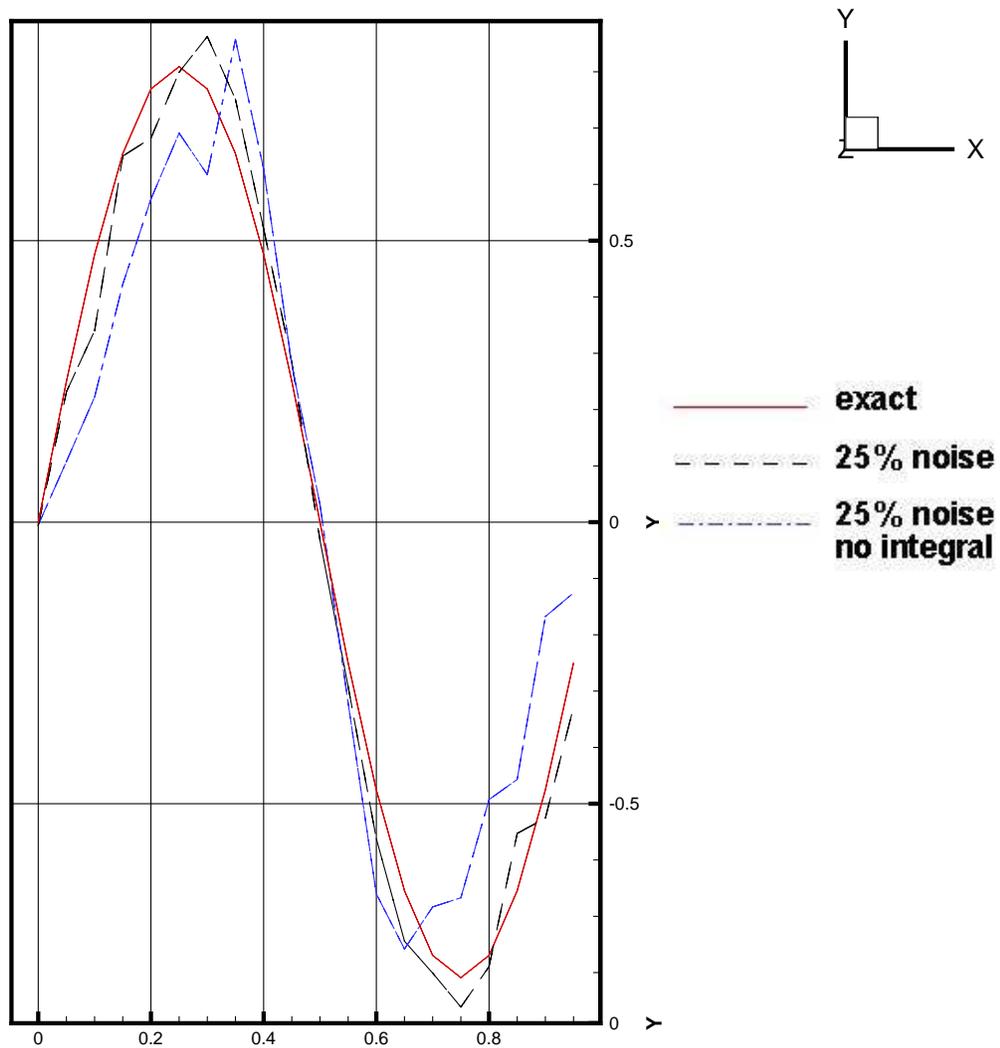}
\end{center}
\caption{Test 3. Cross sections of exact (red) and calculated (black, blue)
functions $\protect\varphi $ with 25\% noise in the boundary data for $%
T=0.75 $, ''no integral'' means $\protect\chi _{\protect\varphi }=0$. The
maximal value of the exact function is $0.9<1$ only because of the grid step
size.}
\label{fig:phi_sin_e_t_25_cross}
\end{figure}

\textbf{Test 3}. \emph{The }$\varphi -$\emph{problem in }$SQ(1)$\emph{\ for }%
$T\in \left( 0.5diam\,SQ\left( 1\right) ,diam\,SQ\left( 1\right) \right) ,$
\emph{where }$diam\,SQ\left( 1\right) =\sqrt{2}$\emph{\ is the diameter of
the square }$SQ\left( 1\right) $. We have decided to see what kind of
results can be obtained if the boundary Cauchy data are given on the entire
boundary of the square $SQ\left( 1\right) $ in the case when $\,T\in \left(
0.5diam\,SQ\left( 1\right) ,diam\,SQ\left( 1\right) \right) .$ We are
especially interested in the question about the influence of terms with $%
\chi _{\varphi }$ and $\chi _{\psi }.$ We have used
\begin{equation*}
\,T=0.75<diam\,SQ\left( 1\right) =\sqrt{2},N_{x}=N_{y}=20,\,N_{t}=3.
\end{equation*}
and have reconstructed the function (7.1). Figure $\ref{fig:phi_sin_e_t_25}$
displays the resulting image with $25\%$ noise in the case when the term $%
\chi _{\varphi }$ is present in (3.1). This quality of the reconstruction is
good for such a high noise level. Figure $\ref{fig:phi_sin_e_t_25_cross}$
displays the 1-dimensional cross-section of the image by the straight line $%
\left\{ x_{1}=0.5\right\} $, as well as the 1-dimensional cross-section of
the image for the case when the term with $\chi _{\varphi }$ is not present
in (3.1) and 25\% noise in the data is in. One can observe that the minimal
value of $(-0.9)$ is not achieved in the case when the term with $\chi
_{\varphi }$ is not present. The calculated minimal value is $(-0.7)$ in
this case.

\begin{figure}[tbp]
\begin{center}
\includegraphics[width=16cm]{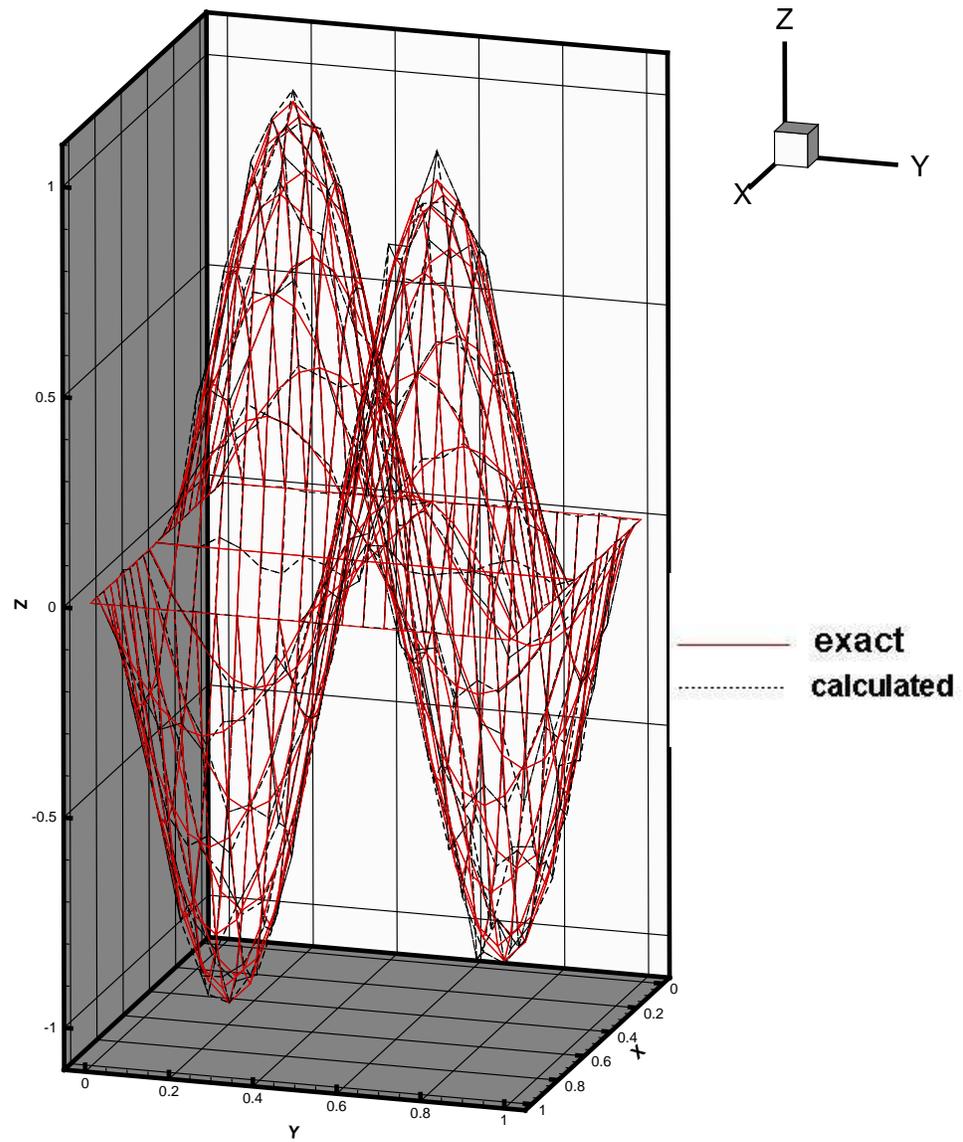}
\end{center}
\caption{Test 4. Exact (red) and calculated (black) solutions of the $%
\protect\varphi -$ problem in SQ(1) with 25\% noise in the boundary data for
$T=2$.}
\label{fig:phi_sin_e_t1_25}
\end{figure}

\begin{figure}[tbp]
\begin{center}
\includegraphics[width=16cm]{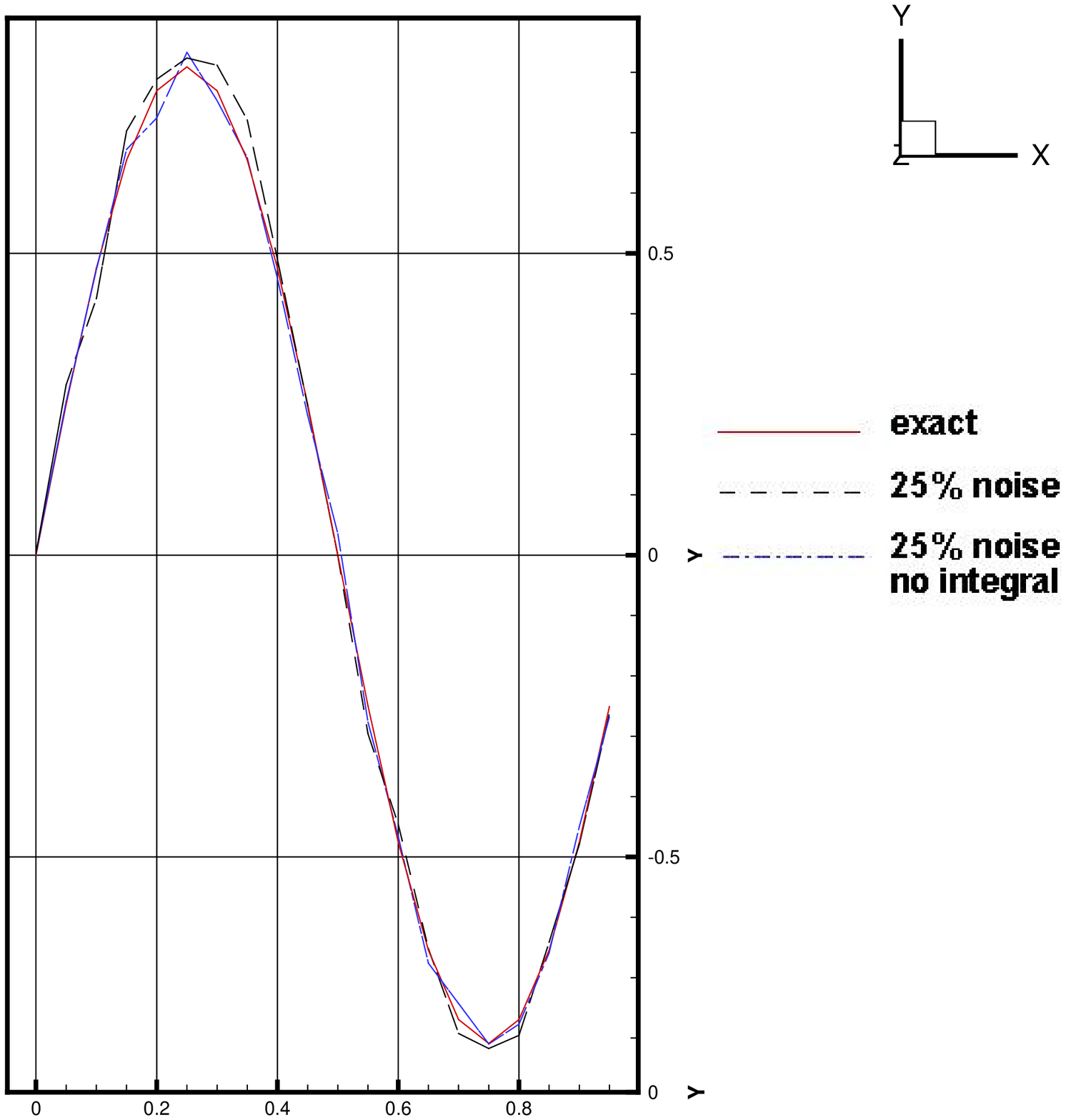}
\end{center}
\caption{Test 4. Cross sections of exact (red) and calculated (black, blue)
functions $\protect\varphi $ with 25\% noise in the boundary data for $%
T=0.75 $, ''no integral'' means $\protect\chi _{\protect\varphi }=0$. The
maximal value of the exact function is $0.9<1$ only because of the grid step
size.}
\label{fig:phi_sin_e_t1_25_cross}
\end{figure}

\textbf{Test 4}. \emph{The }$\varphi -$\emph{problem in $SQ(1)$ for }$%
T>diam\,SQ\left( 1\right) $. We now test our method for the case when the
boundary Cauchy data are given at the entire boundary of the rectangle $%
SQ\left( 1\right) $ and $T>diam\,SQ\left( 1\right) $. We take
\begin{equation*}
T=2,\,N_{x}=N_{y}=20,\,N_{t}=60.
\end{equation*}
The function (7.1) was reconstructed. Figure $\ref{fig:phi_sin_e_t1_25}$
displays the resulting image with $25\%$ noise and Figure $\ref
{fig:phi_sin_e_t1_25_cross}$ displays the 1-dimensional cross-section of the
image by the straight line $\left\{ x_{1}=0.5\right\} $, as well as the
1-dimensional cross-section of the image for the case when the term with $%
\chi _{\varphi }$ is not present in (3.1) (with 25\% noise). One can observe
that both images are very close to the correct one. This points towards the
fact, which follows from the theory of above cited publications and also
from Theorem 4.1: the presence of terms with $\chi _{\varphi }$ and $\chi
_{\psi }$ is important only when $T\in \left( R,2R\right) $ and it is
unimportant for $T>2R.$
\begin{figure}[tbp]
\begin{center}
\includegraphics[width=16cm]{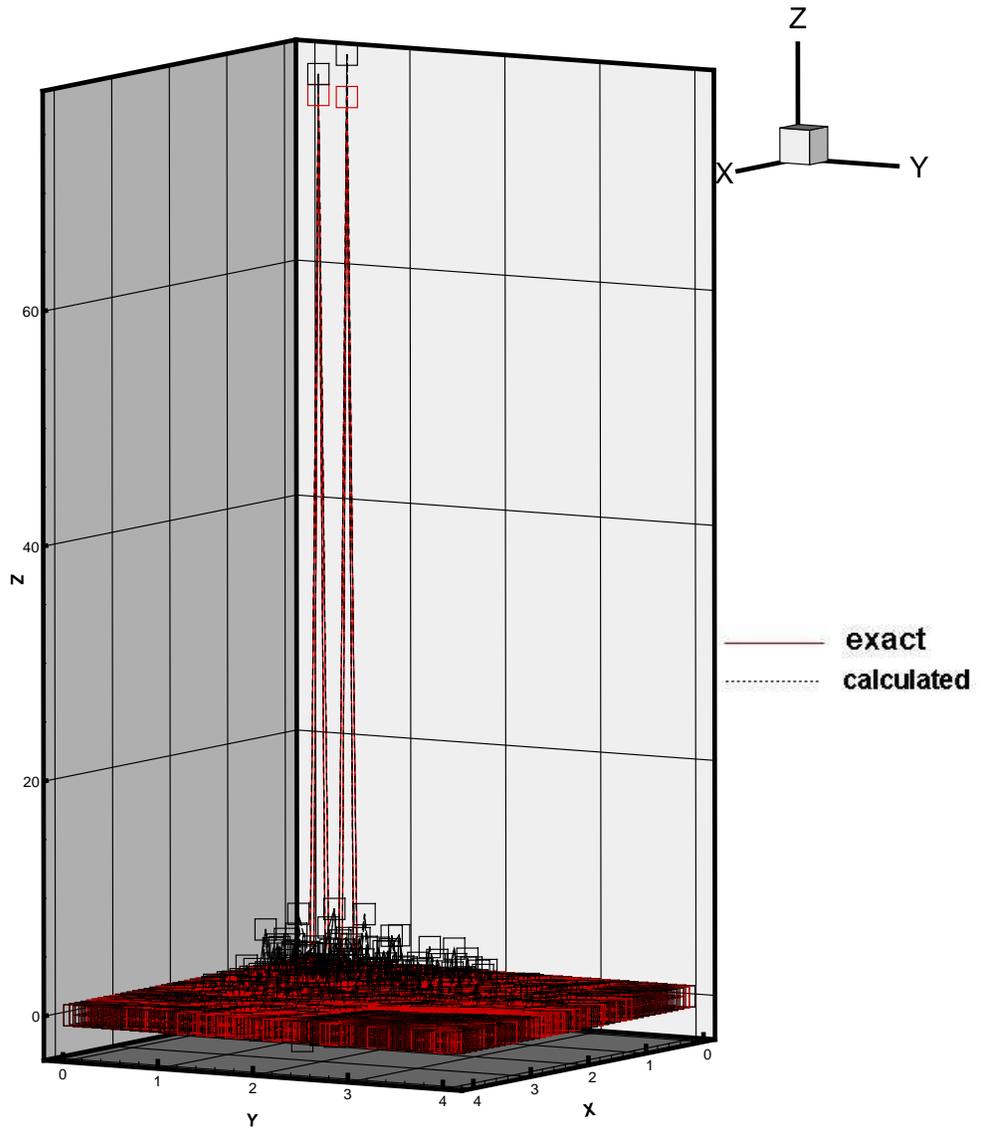}
\end{center}
\caption{Test 5. Exact (red) and calculated (black) function $\protect%
\varphi $ with 50\% noise in the boundary data. The function $\protect\chi _{%
\protect\varphi }$ in (3.1) is present. Scatter plot mode. Squares show
heights. Correct heights are achieved.}
\label{fig:phi_delta_50}
\end{figure}
\begin{figure}[tbp]
\begin{center}
\includegraphics[width=16cm]{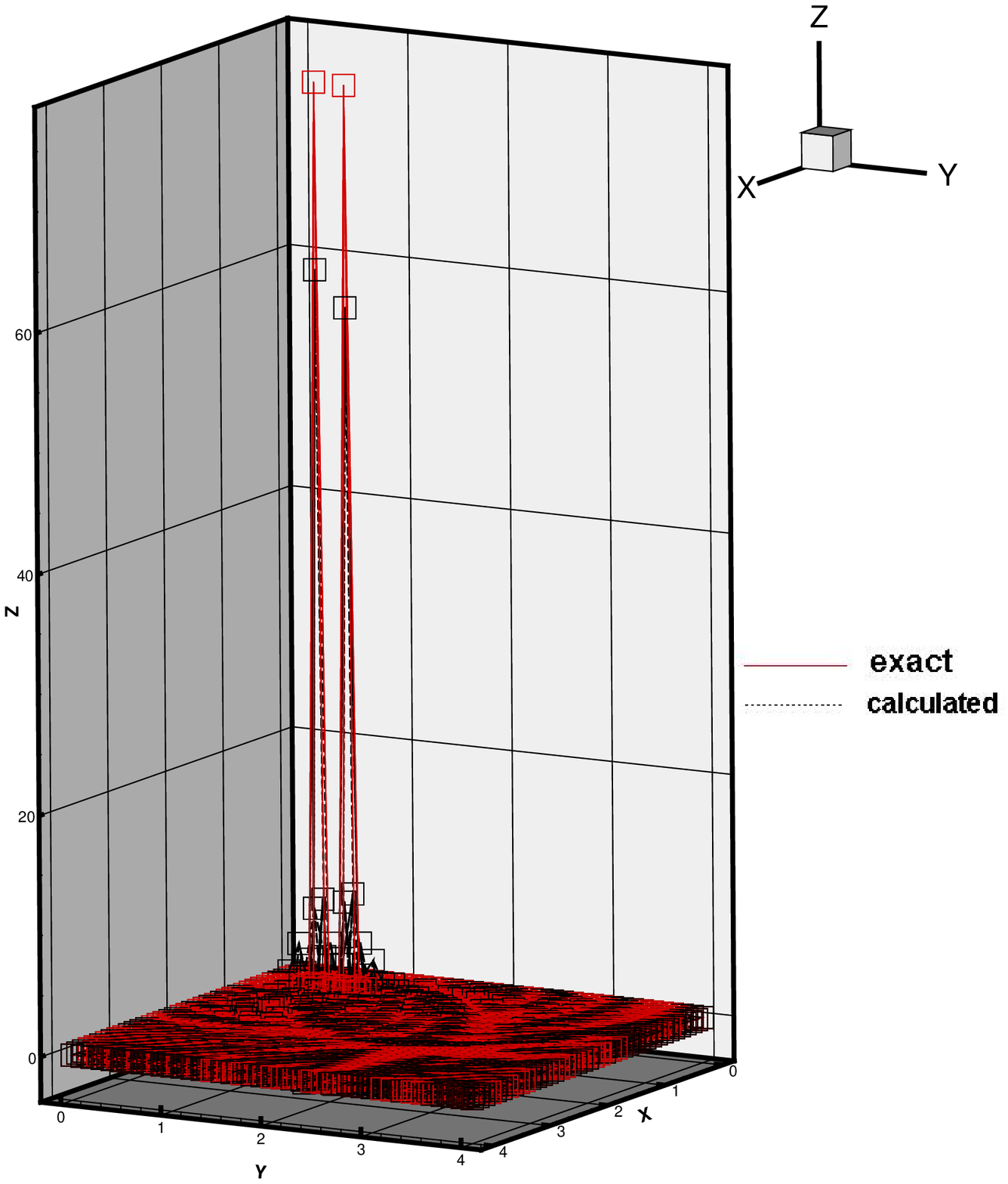}
\end{center}
\caption{Test 5. Exact (red) and calculated (black) function $\protect%
\varphi $ with 5\% noise in the boundary data and $\protect\chi _{\protect%
\varphi }=0$. Scatter plot mode. Squares show heights. Correct heights are
not achieved.}
\label{fig:phi_delta_5_nogt}
\end{figure}

\textbf{Test 5.} \emph{The }$\varphi -$\emph{problem with two }$\delta -$%
\emph{functions. }We now again consider the Inverse Problem 2 with the
domain $\Omega $ as in (6.3) and with $\psi (x)\equiv 0.$\ The data for the
forward problem were simulated for the case
\begin{equation}
\varphi \left( x_{1},x_{2}\right) =\delta \left( x_{1}-0.4,x_{2}-0.4\right)
+\delta \left( x_{1}-0.7,x_{2}-0.7\right)  \tag{7.3}
\end{equation}
with the above described finite difference analogue of the $\delta -$
function. Figure $\ref{fig:phi_delta_50}$ displays the resulting image of
the function (7.3) for the case of 50\% of the noise in the boundary data,
scatter plot mode was used, squares show exact height. Figure $\ref
{fig:phi_delta_5_nogt}$, on the other hand, shows the image when the term
with $\chi _{\varphi }$ is absent in (3.1) and only 5\% noise in the data is
present. One can see that the correct height is not reached on Figure $\ref
{fig:phi_delta_5_nogt}$, unlike Figure $\ref{fig:phi_delta_50}$. This again
shows the importance of the introduction of terms in the third line of (3.1).

Very similar results (not shown) were obtained for the $\psi -$problem with
exactly the same $\delta -$ functions as ones in (7.3).

\section{Conclusions}

We have considered the inverse problems of the determination of one of
initial condition in a hyperbolic equation using the lateral Cauchy data.\
We have presented applications of these problems to the thermoacoustic
tomography, as well as to linearized inverse acoustic and inverse
electromagnetic problems. The problems we consider are very close ones with
the Cauchy problems for hyperbolic equations with the lateral data, and we
have actually solved the latter numerically in Tests 3 and 4. We have
focused on the inverse problem in an infinite domain (octant), whereas only
finite domains were considered in previous numerical studies. Nevertheless,
we are able to reduce our inverse problem to one in a finite domain due to
the finite speed of propagation of waves. Since one initial condition is
known, we were able to decrease the observation time $T$ by twofold. We have
shown numerically that it is important to know one of initial conditions if $%
T<diameter\left( \Omega \right) ,$ as it is required by stability and
uniqueness results. However, if $T>diameter\left( \Omega \right) ,$ then
both the theory and our numerical result of Test 4 show that one does not
need to know the initial condition.

We have proposed a new version of the Quasi-Reversibility method. The main
new element of this version is the inclusion of the terms characterizing
\emph{a priori} knowledge of one of initial conditions. Two other new
elements are incorporation of boundary terms in the Tikhonov functional
instead of subtracting off boundary conditions and the use of finite
differences instead of finite elements in the inverse solver. To prove
convergence of this new version, we have modified the technique of previous
works, which is based on Carleman estimates. A comprehensive numerical study
of the proposed numerical method was conducted. This study has demonstrated
robustness of our technique with respect up to 50\% random noise in the
data, similarly with previous publications [4], [12], [15]. This study has
also demonstrated that this method is capable to image sharp peaks, which is
important for the application to thermoacoustic tomography, for example.

\begin{center}
\textbf{Acknowledgment}
\end{center}

The research of M.V. Klibanov and A. V. Kuzhuget was supported by the U.S.
Army Research Laboratory and U.S. Army Research Office under contract/grant
number W911NF-05-1-0378. The first author has performed a part of this work
during the Special Semester on Quantative Biology Analyzed by Mathematical
Methods, October 1$^{\text{st}}$, 2007 - January 27$^{\text{th}}$, 2008,
organized by RICAM, Austrian Academy of Sciences.

\end{document}